\documentclass[12pt]{article}
\usepackage{graphicx}
\usepackage{amssymb}
\usepackage{alltt}
\usepackage{subfigure}
\usepackage{lineno}

\newcommand{\eg}{e.g., }

\newcommand{\etal}{et al.}
\newcommand{\ie}{i.e., }

\newcommand{\sect}[1]{Section \ref{s:#1}}

\newcommand{\eqn}[1]{Eq.\ (\ref{e:#1})}

\newcommand{\Eqns}[2]{Equations (\ref{e:#1})--(\ref{e:#2})}
\newcommand{\fig}[1]{Fig.\ \ref{f:#1}}

\newcommand{\Fig}[1]{Figure \ref{f:#1}}
\newcommand{\Figtwo}[2]{Figures \ref{f:#1} and \ref{f:#2}}

\newcommand{\tbl}[1]{Table \ref{t:#1}}

\newcommand{\code}[1]{\texttt{#1}} 

\newcommand{\bm}[1]{\mbox{\boldmath{$#1$}}}  

\def\paper#1 #2 #3 #4 #5 #6 {#1, #2. #3. #4\ #5, #6.}
\def\paperL#1 #2 #3 #4 #5 #6 {#1, #2. #3 #4\ #5, #6.}
\def\paperC#1 #2 #3 #4 {#1, #2. #3. #4.}
\def\paperS#1 #2 #3 #4 #5 {#1, #2. #3. #4. #5.}
\def\inpress#1 #2 #3 #4 {#1, #2. #3. #4, in press.}
\def\preprint#1 #2 #3 #4 {#1, #2. #3. ArXiv e-prints #4.}
\def\submitted#1 #2 #3 #4 {#1, #2. #3. #4, submitted.}
\def\inprep#1 #2 #3 #4 {#1, #2. #3. #4, in preparation.}
\def\thesis#1 #2 #3 #4 #5 {#1, #2. #3. Thesis, #4. #5 pp.}
\def\book#1 #2 #3 #4 {#1, #2. #3. #4.}
\def\chap#1 #2 #3 #4 #5 #6 #7 {#1, #2. #3. In: #4 (Eds.), #5. #6, pp.\ #7.}
\def\aIII#1 #2 #3 {#1, 2002. #2. In: Bottke Jr., W.F., Cellino, A.,
  Paolicchi, P., Binzel, R.P. (Eds.), Asteroids III. Univ.\ of Arizona
  Press, Tucson, pp.\ #3.}

\newcommand{\hide}[1]{} 

\newcommand{\figcap}[1]{\item{\bfseries Figure \ref{f:#1}:}}

\renewcommand{\baselinestretch}{2} 
\begin{document} \linenumbers
%
%
\setlength{\footskip}{0pt} 
\begin{center}
  \vspace*{\fill}
  \textbf{\large Numerical predictions of surface effects during the 2029 close approach of asteroid 99942 Apophis}\\
  \vfill
  \textbf{Yang Yu}$^{\star\dag}$\\
  \textbf{Derek C. Richardson}$^\dag$\\
  \textbf{Patrick Michel}$^\ddag$\\
  \textbf{Stephen R. Schwartz}$^{\ddag\dag}$\\
  \textbf{Ronald-Louis Ballouz}$^\dag$\\
  \bigskip
  $^\star$School of Aerospace\\
  Tsinghua University\\
  Beijing 100084 China\\
  \bigskip
  $^\dag$Department of Astronomy\\
  University of Maryland\\
  College Park MD 20740-2421 United States\\
  \bigskip
  $^\ddag$Lagrange Laboratory\\
  University of Nice Sophia Antipolis, CNRS\\
  Observatoire de la C$ˆo$te d’Azur, C.S. 34229, 06304 Nice Cedex 4, France\\
  \bigskip
  Printed \today\\
  \bigskip
  Submitted to \textit{Icarus}\\
  \vfill
  61 manuscript pages\\
  14 figures including 11 in color (online version only)
  \vfill
\end{center}

\newpage
\begin{flushleft}
  Proposed running page head: Surface effects on 99942 Apophis\\
  \bigskip
  Please address all editorial correspondence and proofs to: Derek C. Richardson\\
  \bigskip
  Derek C. Richardson\\
  Department of Astronomy\\
  Computer and Space Sciences Building\\
  Stadium Drive\\
  University of Maryland\\
  College Park MD 20742-2421\\
  Tel: 301-405-8786\\
  Fax: 301-314-9067\\
  E-mail: \code{dcr@astro.umd.edu}
\end{flushleft}

\newpage
\section*{Abstract}

Asteroid (99942) Apophis' close approach in 2029 will be one of the most significant small-body encounter events in the near future and offers a good opportunity for $in$ $situ$ exploration to determine the asteroid's surface properties and measure any tidal effects that might alter its regolith configuration. Resurfacing mechanics has become a new focus for asteroid researchers due to its important implications for interpreting surface observations, including space weathering effects. This paper provides a prediction for the tidal effects during the 2029 encounter, with an emphasis on whether surface refreshing due to regolith movement will occur. The potential shape modification of the object due to the tidal encounter is first confirmed to be negligibly small with systematic simulations, thus only the external perturbations are taken into account for this work (despite this, seismic shaking induced by shifting blocks might still play a weak role and we will look into this mechanism in future work). A two-stage approach is developed to model the responses of asteroid surface particles (the regolith) based on the soft-sphere implementation of the parallel $N$-body gravity tree code \code{pkdgrav}. A full-body model of Apophis is sent past the Earth on the predicted trajectory to generate the data of all forces acting at a target point on the surface. A sandpile constructed in the local frame is then used to approximate the regolith materials; all the forces the sandpile feels during the encounter are imposed as external perturbations to mimic the regolith's behavior in the full scenario. The local mechanical environment on the asteroid surface is represented in detail, leading to an estimation of the change in global surface environment due to the encounter. Typical patterns of perturbation are presented that depend on the asteroid orientation and sense of rotation at perigee. We find that catastrophic avalanches of regolith materials may not occur during the 2029 encounter due to the small level of tidal perturbation, although slight landslides might still be triggered in positions where a sandpile's structure is weak. Simulations are performed at different locations on Apophis' surface and with different body- and spin-axis orientations; the results show that the small-scale avalanches are widely distributed and manifest independently of the asteroid orientation and the sandpile location. We also include simulation results of much closer encounters of the Apophis with Earth than what is predicted to occur in 2029, showing that much more drastic resurfacing takes place in these cases. 

\begin{description}
\item{\textbf{Keywords}: Asteroids, dynamics; asteroids, surfaces; near-Earth objects; regoliths; tides, solid body}
\end{description}

%
%
\newpage
\section{INTRODUCTION} \label{s:intro}

In this paper, we investigate numerically the behavior of granular material at the surface of an asteroid during close approach to the Earth. We focus on the specific case of Asteroid (99942) Apophis, which will come as close as $5.6$ Earth radii on April 13th, 2029.  We provide predictions about possible reshaping and spin-alteration of---and surface effects on---Apophis during this passage, as a function of plausible properties of the constituent granular material. Studies of possible future space missions to Apophis are underway, including one by the French space agency CNES calling for international partners (\eg Michel \etal\ 2012), with the aim of observing this asteroid during the 2029 close encounter and characterizing whether reshaping, spin-alteration, and/or surface motion occur. The numerical investigations presented here allow for estimation of the surface properties that could lead to any observed motion (or absence of motion) during the actual encounter.

Apophis made a passage to the Earth at $\sim 2300$ Earth radii in early 2013.  At that time, the Herschel space telescope was used to refine the determination of the asteroid's albedo and size (M\"uller \etal\ 2013).  According to these observations, the albedo is estimated to be about 0.23 and the longest dimension about $325 \pm 15$ m, which is somewhat larger than previous estimates ($270 \pm 60$ m, according to Delbo' \etal\ 2007). Concurrent radar observations improved the astrometry of the asteroid, ruling out the possibility of a collision with the Earth in 2036 to better than 1 part in $10^6$. However, Wlodarczyk \etal\ (2013) presented a possible path of risk for 2068. This finding has put off any crisis by $\sim 32$ years and makes exploring Apophis in 2029 to be more for scientific interest.  To date, nothing is known about the asteroid's surface mechanical properties, and this is why its close passage in 2029 offers a great opportunity to visit it with a spacecraft, determine its surface properties, and, for the first time, observe potential modifications of the surface due to tidal effects. And as Apophis approaches, it is likely that international interest in a possible mission will increase, since such close approaches of a large object are relatively rare. 

The case for tidally induced resurfacing was made by Binzel \etal\ (2010; also see DeMeo \etal\ 2013) and discussed by Nesvorn\'y \etal\ (2010) to explain the spectral properties of near-Earth asteroids (NEAs) belonging to the Q taxonomic type, which appear to have fresh (unweathered) surface colors. Dynamical studies of these objects found that those bodies had a greater tendency to come close to the Earth, within the Earth-Moon distance, than bodies of other classes in the past $500$ kyr. The authors speculated that tidal effects during these passages could be at the origin of surface material disturbance leading to the renewed exposure of unweathered material. We leave a more general and detailed investigation of this issue for future work, but if this result is true for those asteroids, it may also be true for Apophis, which will approach Earth on Friday, April 13, 2029 no closer than about 29,500 km from the surface (\ie $4.6$ Earth radii, or $5.6$ Earth radii from the center of the planet; Giorgini \etal\ 2008).  It is predicted to go over the mid-Atlantic, appearing to the naked eye as a moderately bright point of light moving rapidly across the sky.  Our aim is to determine whether, depending on assumed mechanical properties, it could experience surface particulate motions, reshaping, or spin-state alteration due to tidal forces caused by Earth's gravity field. The classical Roche limit for a cohesionless fluid body of bulk density $2.4$ $\mathrm{g/cm^3}$ to not be disrupted by tidal forces is $\sim 3.22$ Earth radii, so we do not expect any violent events to occur during the rocky asteroid's 2029 encounter at $5.6$ Earth radii.

The presence of granular material (or regolith) and boulders at the surface of small bodies has been demonstrated by space missions that visited or flew by asteroids in the last few decades (\eg Veverka \etal\ 2000; Fujiwara \etal\ 2006). It appears that all encountered asteroids to date, from the largest one, the main belt Asteroid (4) Vesta by the Dawn mission, to the smallest one, the NEA (25143) Itokawa, sampled by the Hayabusa mission, are covered with some sort of regolith. In fact, thermal infrared observations support the idea that most asteroids are covered with regolith, given their preferentially low thermal inertia (Delbo' \etal\ 2007).  There even seems to be a trend as a function of the asteroid's size based on thermal inertia measurements: larger objects are expected to have a surface covered by a layer of fine regolith, while smaller ones are expected to have a surface covered by a layer of coarse regolith (Clark \etal\ 2002).  This trend is consistent with observations by the NEAR-Shoemaker spacecraft of the larger ($\sim 17$ km mean diameter) Eros, whose surface is covered by a deep layer of very fine grains, and by the Hayabusa spacecraft of the much smaller ($320$ m mean diameter) Itokawa, whose surface is covered by a thin layer of coarse grains.  However, interpretation of thermal inertia measurements must be made with caution, as we do not yet have enough comparisons with actual asteroid surfaces to verify that the suggested trend is systematically correct. 

Thus, we are left with a large parameter space to investigate possible surface motion during an Earth close approach of an asteroid with unknown surface mechanical properties.  Our approach is to consider a range of simple and well-controlled cases that certainly do not cover all possibilities regarding Apophis' surface mechanical properties, but rather aim at demonstrating whether, even in a simple and possibly favorable case for surface motion, some resurfacing event can be expected to occur during the passage. For instance, instead of considering a flat granular surface, we consider a sandpile consisting of a size distribution of spherical grains (\sect{initialcondition}) and vary the grain properties in order to include more or less favorable cases for motion (from a fluid-like case to a case involving rough particles). Slight disturbances may manifest as very-small-scale avalanches in which grain connections readjust slightly, for example. The forces acting on the sandpile are obtained by measuring all ``external'' perturbations during the encounter, including body spin magnitude and orientation changes, for cases in which the global shape remains nearly fixed, and again assuming simple and favorable configurations of the asteroid. Indirectly, the encounter may also lead to internal reconfigurations of the asteroid, which in turn produce seismic vibrations that could propagate to the surface and affect the regolith material. These secondary modifications are not modeled here, although it may be possible in future work to account for this by shaking the surface in a prescribed manner. In any case, for this particular encounter, we demonstrate (\sect{reshpeff}) that any global reconfiguration will likely be small to negligible in magnitude.

In the following, we first present, in \sect{method}, the numerical method used to perform our investigation, including the initial conditions of the sandpile adopted to investigate surface motion, the representation of the encounter, and the mechanical environment. Results are described in \sect{results}, including potential reshaping of the asteroid, tidal disturbances for Apophis' encounter in 2029, which is a function of the sandpile properties, spin orientation changes, and the dependency of the location of the sandpile on the asteroid to the outcome of the encounter. We also show the responses of the sandpiles for artificially close approaches ($4.0$ and $2.0$ Earth radii) to demonstrate that our method does predict significant alteration of the sandpiles when this is certainly expected to happen. The investigation is discussed in \sect{discuss} and conclusions are presented in \sect{concl}.

\section{NUMERICAL METHOD} \label{s:method}

We use \code{pkdgrav}, a parallel $N$-body gravity tree code (Stadel
2001) adapted for particle collisions (Richardson \etal\ 2000; 2009;
2011).  Originally collisions in \code{pkdgrav} were treated as
idealized single-point-of-contact impacts between rigid spheres.  A
soft-sphere option was added recently (Schwartz \etal\ 2012); with
this new functionality, particle contacts last many timesteps, with reaction
forces dependent on the degree of overlap (a proxy for surface
deformation) and contact history---this is appropriate for dense
and/or near-static granular systems with multiple persistent contacts per particle.  The code uses a 2nd-order leapfrog integrator, with accelerations due to gravity and contact
forces recomputed each step.  Various types of user-definable
confining walls are available that can be combined to provide complex
boundary conditions for the simulations.  The code also includes an
optional variable gravity field based on a user-specified set of
rules.

The spring/dash-pot model used in \code{pkdgrav}'s soft-sphere
implementation is described fully in Schwartz \etal\ (2012).  Briefly,
a (spherical) particle overlapping with a neighbor or confining wall
feels a reaction force in the normal and tangential directions
determined by spring constants ($k_n$, $k_t$), with optional damping
and effects that impose static, rolling, and/or twisting friction.
The damping parameters ($C_n$, $C_t$) are related to the conventional
normal and tangential coefficients of restitution used in hard-sphere
implementations, $\varepsilon_n$ and $\varepsilon_t$.  The static,
rolling, and twisting friction components are parameterized by
dimensionless coefficients $\mu_s$, $\mu_r$, and $\mu_t$,
respectively. Plausible values for these parameters are
obtained through comparison with laboratory experiments (also see \sect{initialcondition}).  Careful consideration of the soft-sphere parameters is needed to ensure internal consistency, particularly with the choice of $k_n$, $k_t$, and timestep---a separate code is provided to assist the user with configuring these parameters correctly.  The numerical approach has been validated through comparison with laboratory experiments; \eg
Schwartz \etal\ (2012) demonstrated that \code{pkdgrav} correctly
reproduces experiments of granular flow through cylindrical hoppers,
specifically the flow rate as a function of aperture size, and found the material properties of the grains also affect the flow rate. Also simulated successfully with the soft-sphere code in \code{pkdgrav} were laboratory impact experiments into sintered glass beads (Schwartz \etal\ 2013), and regolith, in support of asteroid sampling mechanism design (Schwartz \etal\ 2014). 

We use a two-stage approach to model the effect of a tidal encounter on asteroid surface particles (regolith).  First, a rigid (non-deformable) object approximating the size, shape, and rotation state of the asteroid is sent past the Earth on a fixed trajectory (in the present study, the trajectory is that expected of (99942) Apophis---see \sect{representME}; note that the actual shape of Apophis is poorly known beyond an estimate of axis ratios, so we assume an idealized ellipsoid for this study). All forces acting at a target point designated on the object surface are recorded (\sect{representME}).  Then, a second simulation is performed in the local frame of the target point, allowing the recorded external forces to affect the motion of particles arranged on the surface (in the present study we consider equilibrated sandpiles).  This two-stage approach is necessary due to the large difference in scale between the asteroid as a whole and the tiny regolith particles whose reactive motion we are attempting to observe. We approximate the regolith, which in reality likely consists of a mixture of powders and gravel (Clark \etal\ 2002), by a size distribution of solid spheres.  We mimic the properties of different materials by adjusting the soft-sphere parameters (\sect{initialcondition}). The soft-sphere approach permits simulation of the behavior of granular materials in the near-static regime, appropriate for the present case in which the regolith particles generally remain stationary for hours and suffer a rapid disturbance only during the moments of closest approach to the planet. In particular, the model permits a detailed look at the responses of local individual particles even when the tidal effects are too weak to cause any macroscopic surface or shape changes; this gives insight into the limit of tidal resurfacing effects.

\subsection{Sandpile Initial Conditions} \label{s:initialcondition}

In order to assess the effect of tidal encounters on a small surface
feature, we carried out numerical simulations in a local frame
consisting of a flat horizontal surface (i.e., the local plane tangential to the asteroid surface at the target point) with a ``sandpile'' resting on top.  The sandpile consists of $N$ = 1683 simulated spherical particles with radii drawn from a power-law distribution of mean $1.46$ cm and $\pm 0.29$ cm width truncated (minimum and maximum particle radii $1.17$ and $1.75$ cm, respectively). Between the sandpile and the floor is a rigid monolayer of particles drawn from the same distribution and laid down in a close-packed configuration in the shape of a flat disk.  This rigid particle layer provides a rough surface for the sandpile to reduce spreading (\fig{sandpile}). Three different sandpiles were constructed using the material parameters described below.  Particles were dropped though a funnel from a low height onto the rough surface to build up the sandpile. These sandpiles were allowed to settle until the system kinetic energy dropped to near zero. This approach eliminates any bias that might arise from simply arranging the spheres by hand in a geometrical way; the result should better represent a natural sandpile.

\begin{center}
\Fig{sandpile}.
\end{center}

For this study, we compared three different sets of soft-sphere parameters for
the sandpile particles (\tbl{SSDEMparams}).  Our goal was to define a
set of parameters that spans a plausible range of material properties
given that the actual mechanical properties of asteroid surface
material are poorly constrained.  In the specific case of Apophis,
very little is known beyond the spectral type, Sq (Binzel
\etal\ 2009).  There are no measurements of thermal properties that
might give an indication of the presence or absence of regolith on
Apophis.  Consequently, we chose three sets of material parameters
that span a broad range of material properties.  The first set,
denoted ``smooth'' in the table, consists of idealized frictionless
spheres with a small amount of dissipation (5\%, chosen to match the
glass beads case).  This is about as close to the fluid case that a sandpile
can achieve while still exhibiting shear strength arising from the
discrete nature of the particles (and the confining pressure of
surface gravity). It is assumed this set will respond most readily
to tidal effects due to the absence of friction between the particles and the non-uniform size distribution. The second set, ``glass beads,'' is modeled after actual glass beads
being used in a set of laboratory experiments to calibrate numerical
simulations of granular avalanches (Richardson \etal\ 2012).  In this
case $\varepsilon_t$ was measured directly, which informed our choice for $C_t$ (Schwartz \etal\ 2012), and $\mu_s$ and $\mu_r$ were inferred from matching the simulations to the experiments.  The glass beads provide an intermediate case between the near-fluid smooth spheres and the third parameter set, denoted ``gravel.''

\begin{center}
\tbl{SSDEMparams}.
\end{center}

The gravel parameters were arrived at by carrying out simple avalanche
experiments using roughly equal-size rocks collected from a streambed.  In these experiments, the rocks (without sharp edges) were released near the top of a wooden board held at a 45$^\circ$ incline.  The dispersal pattern on the stone floor was measured, including the distance from the end of the board to approximately half of the
material, the furthest distance traveled by a rock along the direction
of the board, and the maximum angle of the dispersal pattern relative
to the end of the board (\fig{rockslide}a--c).  A series of numerical simulations was then performed to reproduce the typical behavior by varying the soft-sphere
parameters (\fig{rockslide}d--f).  The values used in \tbl{SSDEMparams} for ``gravel'' were found in this way.  The $\mu_s$ and $\mu_r$ values are quite large,
reflecting the fact that the actual particles being modeled were not spheres.  A
correspondingly smaller timestep is needed to adequately sample the
resulting forces on the particles in the simulations.  The value of
$\varepsilon_n$ was measured by calculating the average first rebound height of sample gravel pieces that were released from a certain height; $\varepsilon_t$ was not measured
but since the rocks were rough it was decided to simply set
$\varepsilon_t = \varepsilon_n$.  This exercise was not meant to be an
exhaustive or precise study; rather, we sought simply to find
representative soft-sphere parameters that can account plausibly for the
irregularities in the particle shapes.  In any case, we will find that
such rough particles are difficult to displace using tidal forces in
the parameter range explored here, so they provide a suitable upper
limit to the effects.  Similarly, we do not consider cohesion in this
numerical study, which would further resist particle displacement due
to tidal forces.

\begin{center}
\Fig{rockslide}.
\end{center}

\subsection{Representation of Mechanical Environment} \label{s:representME}

A new adaptation of \code{pkdgrav} developed in this work allows for the simulation of sandpiles located on an asteroid surface, based on the motion equations of granular material in the noninertial frame fixed to the chosen spot. Detailed mechanics involved in the local motion of the sandpile during a tidal encounter were considered and represented in the code, including the contact forces (reaction, damping, and friction between particles and/or walls) and the inertial forces derived from an analysis of the transport motion of a flyby simulation. Here we provide a thorough derivation of the relevant external force expressions.

\begin{center}
\Fig{frames}.
\end{center}

\Fig{frames} illustrates four frames used in the derivation, the space inertial frame (SPC), mass center translating frame (MCT), body fixed frame (BDY) and local frame (LOC). \eqn{PosSpc2Loc} gives the connection between the spatial position and the local position of an arbitrary particle in the sandpile, for which $\mathbf{R}$ is the vector from SPC's origin to the particle, $\mathbf{R}_C$ is the vector from SPC to MCT/BDY, $\mathbf{l}$ is the vector from MCT/BDY to LOC, and $\mathbf{r}$ is the vector from LOC to the particle: 

\begin{equation} 
\label{e:PosSpc2Loc}
\mathbf{R}=\mathbf{R}_C+\mathbf{l}+\mathbf{r}.
\end{equation}

\Eqns{VelSpc2Loc}{AccSpc2Loc} are derived by calculating the first- and second-order time derivatives of \eqn{PosSpc2Loc}, which denote the connections of velocity and acceleration between the spatial and local representations, respectively. $\bm \omega$ indicates the angular velocity vector of the asteroid in SPC or MCT. The operator $\frac{\mathrm{d}}{\mathrm{d} t}$ denotes the time derivative with respect to the inertial frame SPC, while $\frac{\tilde \mathrm{d}}{\mathrm{d} t}$ denotes the time derivative with respect to the body-fixed frame BDY or LOC: 

\begin{equation}
\label{e:VelSpc2Loc}
\frac{\mathrm{d} }{\mathrm{d} t} \mathbf{R}=\frac{\mathrm{d} }{\mathrm{d} t} \mathbf{R}_C+{\bm \omega} \times (\mathbf{l}+\mathbf{r})+\frac{\tilde \mathrm{d}}{\mathrm{d} t} \mathbf{r},
\end{equation}

\begin{equation}
\label{e:AccSpc2Loc}
\frac{\mathrm{d}^2 }{\mathrm{d} t^2} \mathbf{R}=\frac{\mathrm{d}^2}{\mathrm{d} t^2} \mathbf{R}_C+{\bm \omega} \times [{\bm \omega} \times (\mathbf{l}+\mathbf{r})]+\frac{\tilde \mathrm{d}}{\mathrm{d} t} {\bm \omega} \times (\mathbf{l}+\mathbf{r})+2 {\bm \omega} \times \frac{\tilde \mathrm{d}}{\mathrm{d} t} \mathbf{r}+\frac{\tilde \mathrm{d}^2 }{\mathrm{d} t^2} \mathbf{r}.
\end{equation}

The dynamical equation for an arbitrary particle in the inertial frame SPC can be written as \eqn{SpcDynEqn}, which shows that the forces a particle in the sandpile feels can be grouped into three categories: $\mathbf{F}_A$ denotes the local gravity from the asteroid; $\mathbf{F}_P$ denotes the attraction from the planet; and $\mathbf{F}_C$ represents all the contact forces coming from other particles and walls:

\begin{equation}
\label{e:SpcDynEqn}
\frac{\mathrm{d}^2 }{\mathrm{d} t^2} \mathbf{R}=\mathbf{F}_A+\mathbf{F}_P+\mathbf{F}_C. 
\end{equation}

We get the corresponding dynamical equation in the local frame LOC (\eqn{LocDynEqu}) by substituting \eqn{AccSpc2Loc} into \eqn{SpcDynEqn},

\begin{equation}
\label{e:LocDynEqu}
\frac{\tilde \mathrm{d}^2 }{\mathrm{d} t^2} \mathbf{r}=\mathbf{F}_C+\mathbf{F}_A+\mathbf{F}_P-\frac{\mathrm{d}^2}{\mathrm{d} t^2} \mathbf{R}_C-{\bm \omega} \times [{\bm \omega} \times (\mathbf{l}+\mathbf{r})]-\frac{\tilde \mathrm{d}}{\mathrm{d} t} {\bm \omega} \times (\mathbf{l}+\mathbf{r})-2 {\bm \omega} \times \frac{\tilde \mathrm{d}}{\mathrm{d} t} \mathbf{r}.
\end{equation}

The terms on the right-hand side of \eqn{LocDynEqu} represent the mechanical environment of a particle in the local sandpile during a tidal encounter, and can be interpreted as follows. $\mathbf{F}_C$ is the reaction force from the walls and surrounding particles, which is directly exported by \code{pkdgrav}. $\mathbf{F}_A$ is approximated by a constant value at the LOC's origin that acts as the uniform gravity felt throughout the whole sandpile, since the latter's size is negligible compared with the asteroid dimensions. The difference between the planetary attraction and the translational transport inertia force, $\mathbf{F}_P-\frac{\mathrm{d}^2}{\mathrm{d} t^2} \mathbf{R}_C$, is the tidal force that plays the primary role of moving the surface materials. The term $-{\bm \omega} \times [{\bm \omega} \times (\mathbf{l}+\mathbf{r})]$ is the centrifugal force due to the rotation of the asteroid, $-\frac{\tilde \mathrm{d}}{\mathrm{d} t} {\bm \omega} \times (\mathbf{l}+\mathbf{r})$ is the librational transport inertia force (LTIF) that results when the rotational state changes abruptly during the encounter, and $-2 {\bm \omega} \times \frac{\tilde \mathrm{d}}{\mathrm{d} t} \mathbf{r}$ is the Coriolis effect that only acts when particle movement occurs. 

\begin{center}
\Fig{rubblepile}.
\end{center}

In the numerical scheme, we derived the component dynamical equations by projecting \eqn{LocDynEqu} to LOC and treating each term separately. The local gravity $\mathbf{F}_A$ and LOC origin $\mathbf{l}$ are constant vectors, which were initialized in the simulations as parameters. The complications arising from the asteroid motion (tidal force, angular velocity, and acceleration) were solved within a simulation in advance, using a rigid rubble pile as the asteroid model and exporting the physical quantities required in each step. To specify Apophis's encounter with Earth in 2029, we computed the planet trajectory by tracing back from the perigee conditions, $5.6$ $\mathrm{R_E}$ distance and $8.4$ km/s encounter speed (Giorgini \etal\ 2008), where $\mathrm{R_E}$ indicates the average Earth radius ($6371$ km). As \fig{rubblepile} illustrates, a tri-axial ellipsoid model with axis ratio 1.4:1.0:0.8 was employed to approximate the overall shape of Apophis (Scheeres \etal\ 2005), corresponding to an equivalent radius of $162.5$ m (M\"uller \etal\ 2013). A rubble pile bounded by the tri-axial model was constructed for the flyby simulation. The rubble pile model consisted of $N$ = 2855 equal-size spherical particles in a close-packed configuration. The total mass was set to $4 \times 10^{10}$ kg (bulk density $2.4$ $\mathrm{g/cm^3}$) with an initial rotation around the maximum principle axis of inertia of period $30.4$ h (Tholen \etal\ 2013). Three markers (non-coplanar with the origin) were chosen from the surface particles to track the variation of the rubble pile's attitude in SPC. A complete list of its motion states during the encounter was exported to a data file, which was accessed by the subsequent simulations with local sandpiles. 

\section{RESULTS} \label{s:results}

\subsection{Evaluation of Reshaping Effects} \label{s:reshpeff}

Richardson \etal\ (1998) showed the complicated behavior of a rotating rubble pile due to a tidal encounter. Generally, a variation in rotational state is usually induced because of the coupling effect between librational motion and orbital motion. Especially for elongated bodies like Apophis, the terrestrial torques during its flyby may force a strong alteration of the rotational state in a short period (Scheeres \etal\ 2005). The modifications in structure of the rubble pile are also notable. The tidal encounter outcomes show a strong dependence on the progenitor elongation, the perigee distance, and the time spent within the Roche sphere. The outcomes also depend on the spin orientation of the progenitor to the extent that retrograde rotators always suffer less catastrophic consequences than prograde ones (Richardson \etal\ 1998). However, the parameter space considered in that study (Richardson \etal\ 1998) did not include the Apophis scenario in 2029, since the focus was on the conditions for full-scale distortion or disruption. For this study, we first redid the flyby simulations (\sect{representME}) using a true (non-rigid) rubble pile model with the Apophis encounter parameters of $5.6$ $\mathrm{R_E}$ and $8.4$ km/s. We quantified the tidal reshaping effect by measuring the maximum net change of a normalized shape factor defined by \eqn{ShpFact} (Hu \etal\ 2004), in which  $I_1$, $I_2$, $I_3$ are the principal inertia moments, with $I_1 \leqslant I_2 \leqslant I_3$:

\begin{equation}
\label{e:ShpFact}
\sigma \equiv \frac{I_2-I_1}{I_3-I_1}.
\end{equation}

The shape factor $\sigma \in [0,1]$ roughly describes the mass distribution of the body (0=oblate, 1=prolate). The initial rubble-pile model (equilibrated, prior to the encounter) had $\sigma \sim 0.73$, and its relative change, defined by $\Delta \sigma / \sigma$, was measured for several simulations parameterized by bulk density and constituent material type (\tbl{FullScaleChange}). The corresponding Roche limits are provided in the table, with the values estimated for the case of a circular orbit. The reshaping effects show consistent results for all three types of materials, varying between a noticeable change in shape (magnitude $10^{-2}$) and a negligible change in shape (magnitude $10^{−5}$), with the sharpest transition occurring for a critical bulk density $\sim 0.4$ $\mathrm{g/cm^3}$ with corresponding Roche limit of $5.6$ $\mathrm{R_E}$. This agrees with the analysis of Apophis' disruption limit using a continuum theory for a body made up of solid constituents (Holsapple \etal\ 2006). Note we find that even a bulk density as low as $0.1$ $\mathrm{g/cm^3}$ did not  dislodge any particles enough for them to end up in a new geometrical arrangement, due to the very short duration of the tidal encounter. 

\begin{center}
\Fig{newpile}.
\end{center}

\begin{center}
\tbl{FullScaleChange}.
\end{center}

Since idealized equal-size spheres can arrange into crystal-like structures that may artificially enhance the shear strength of the body (Tanga \etal\ 2009; Walsh \etal\ 2012), we carried out a second suite of simulations with rubble piles made up of a bimodal distribution of spheres (\fig{newpile}). The model consisted of $181$ big particles of radius $19.3$ m and $1569$ small particles of radius $9.7$ m, bounded by the same tri-axial ellipsoid as shown in \fig{rubblepile}, and arranged randomly. As before, the bulk density and material type were varied, with notable differences in outcome compared to the equal-size-sphere cases (\tbl{FullScaleChange}). Particles using the ``smooth" parameter set are not able to hold the overall shape of Apophis (the 1.4:1.0:0.8 tri-axial ellipsoid); instead the rubble pile collapses and approaches a near-spherical shape. For the other two parameter sets, however, the overall shape of the rubble pile is maintained. \Fig{comp} shows the relative change of shape factor $\sigma$ as a function of bulk density for both types of simulations. As shown, the results from the unimodal and bimodal rubble piles present similar trends: before reaching the Roche limit density, the reshaping remains small (the magnitudes are $10^{-5}$--$10^{-4}$ with a certain amount of stochasticity); after that, the reshaping effect sharply increases to around $10^{-1}$. The bimodal rubble piles show less resistance to the tidal disturbance, since they have less-well-organized crystalline structures compared to the unimodal cases; they therefore exhibit larger shape modifications (Walsh \etal\ 2012).

\begin{center}
\Fig{comp}.
\end{center}

Regardless, catastrophic events were not detected for either type of rubble pile until the bulk density was as small as $0.1$ $\mathrm{g/cm^3}$, for all material parameter sets. Therefore, the reshaping effects on Apophis in 2029 should be negligibly small for bulk densities in the likely range ($2$--$3$ $\mathrm{g/cm^3}$). Even so, minor internal reconfigurations resulting from the tidal encounter may produce seismic waves that could propagate and affect the configuration of surface regolith. Since we are only looking at localized areas on the surface, isolated from the rest of the body, vibrations emanating from other regions, in or on the asteroid, are not evaluated in the current work, although again we expect these to be small or non-existent for the specific case of the Apophis 2029 encounter. Only the ``external" forces acting directly on the surface particles in the considered localized region are taken into account. We will look into the effects of seismic activity, which may stem from other regions of the asteroid, on an actual high-resolution rubble pile in future work. In this study, we focus on external forces, outside of the rubble pile itself; the rigid rubble pile model is employed as a reasonable simplification for the purpose of measuring the external forces on a surface sandpile. 

\subsection{Global Change of Mechanical Environment} \label{s:GCME}

The right-hand side of \eqn{LocDynEqu} provides a description of the constituents of the mechanical environment experienced by a sandpile particle, including local gravity, tidal force, centrifugal force, LTIF, and the Coriolis effect. Note the Coriolis force does not play a role before particle motion begins, so it can be ignored when examining the causes of avalanches/collapses. The centrifugal force and LTIF show weak dependence on local position $\mathbf{r}$, so these terms can be simplified by substituting $\mathbf{l}+\mathbf{r} \approx \mathbf{l}$, since the dimension of the sandpile is much smaller than that of the asteroid. This leads to an approximation of the resultant environmental force (\eqn{EnvirForce}), which provides a uniform expression of the field force felt throughout the sandpile:  

\begin{equation}
\label{e:EnvirForce}
\mathbf{F}_E=\mathbf{F}_A+\mathbf{F}_P-\frac{\mathrm{d}^2}{\mathrm{d} t^2} \mathbf{R}_C-{\bm \omega} \times ({\bm \omega} \times \mathbf{l})-\frac{\tilde \mathrm{d}}{\mathrm{d} t} {\bm \omega} \times \mathbf{l}.
\end{equation}

The variation of $\mathbf{F}_E$ shows a common pattern for different encounter trajectories and different locations on the asteroid: it stays nearly invariable for hours as the planet is approaching/departing and shows a rapid single perturbation around the rendezvous. This feature enables us to evaluate the mechanical environment by measuring the short-term change of $\mathbf{F}_E$, and moreover, to make a connection between these external stimuli with the responses of sandpiles in the simulations. Spherical coordinates $(F, \theta, \alpha)$ are used to represent the resultant force (\fig{ACCS}), in which $F$ is the magnitude and $\theta$ and $\alpha$ denote the effective gravity slope angle and deflection angle, respectively. 

\begin{center}
\Fig{ACCS}.
\end{center}

The effects of the environment force must be considered in the context of the sandpile modeling. A rough approximation of the catastrophic slope angle (\eqn{CritcAng}) can be derived from the conventional theory of the angle of repose for conical piles (Brown \etal\ 1966), namely that avalanches will occur when the effective gravity slope angle $\theta$ exceeds $\theta_{C}$: 

\begin{equation}
\label{e:CritcAng}
\theta_C=\tan^{-1}(\mu_S)-\theta_S. 
\end{equation}

Here $\theta_S$ denotes the resting angle of the (conical) sandpile, assumed to be less than the repose angle $\tan^{-1}(\mu_S)$, and $\theta_C$ denotes the critical slope angle, which, if exceeded by the effective gravity slope angle, results in structural failure of the sandpile, \ie an avalanche. Essentially, the avalanches of the sandpile depend primarily on the instantaneous change of the slope angle $\theta$ of the resultant force and should be only weakly related to the magnitude $F$ and deflection angle $\alpha$. The global distribution of changes of $\theta$ were examined for the duration of the encounter (see below). It would be difficult to make an exhaustive search over all asteroid encounter orientations due to the coupling effects between the asteroid's rotation and the orbital motion. Instead, we chose $12$ representative trajectories along the symmetry axes of the Apophis model (at perigee) for study, which serves as a framework for understanding the influence of encounter orientation and the strength of tidal disturbance at different locations. (The technique used to match the orientation of the asteroid at perigee is to first run a simulation in reverse starting at perigee with the required orientation, thereby providing the correct asteroid state for a starting position far away.) \Fig{traject} shows the $12$ trajectories along three mutually perpendicular planes, including both the prograde and retrograde cases. This choice is based on the symmetry of the dynamical system, and all $12$ trajectories have a speed of  $8.4$ km/s at a perigee distance of $5.6$ $\mathrm{R_E}$. 

\begin{center}
\Fig{traject}.
\end{center}

\Fig{traject} presents these representative trajectories in order, each corresponding to a special possible orientation of the encounter. The effective gravity slope angle changes throughout the surface of the tri-axial ellipsoid model were recorded for each trajectory, with particular attention to the location and time of the maximum change. We found that these maximum changes concentrate at several minutes around perigee. \Fig{maxCSA} shows the maximum values of slope angle change along these trajectories, indicating that the largest perturbation on slope angle is less than $2^\circ$. We verified that the achievable range of the effective gravity slope angle during the encounter (\sect{representME}) is within the safe limit predicted by \eqn{CritcAng} for all three sandpiles (\sect{initialcondition}); that is, a slope change below $2^\circ$ is not enough to trigger any massive avalanches throughout the sandpiles.  

\begin{center}
\Fig{maxCSA}.
\end{center}

\Fig{maxCSA} also shows a general dependence of the tidal perturbation on the orientation of the encounter trajectory, namely that the three step levels seen in the figure correspond to three directions of the relative planet motion at perigee. Trajectories $1$--$4$ correspond to the direction of the body long axis, which leads to relatively strong tidal effects; trajectories $5$--$8$ correspond to the direction of the intermediate axis, which leads to moderate effects; while trajectories $9$--$12$ correspond to the direction of the short axis, which leads to relatively weak tidal effects. \Fig{patterns} illustrates the distribution of effective gravity slope angle change at perigee for the three trajectory sets, with A denoting trajectories $1$--$4$, B denoting trajectories $5$--$8$, and C denoting trajectories $9$--$12$. We confirm that the four trajectories in each set present visually the same distribution around perigee, therefore we chose the patterns due to trajectories 1, 5, and 9 for demonstration of sets A, B, and C, and generalize these three patterns as representative. Several points can be inferred from \fig{patterns}. First, the direction to the planet at perigee largely determines the global distribution of tidal perturbation, that is, the strongest effects tend to occur along the long axis and the weakest effects along the short axis. Second, at perigee, the largest slope change occurs near areas surrounding the pole for the most favorable orientation (set A), while the largest slope change at the pole occurs several minutes before or after perigee (see \fig{pxpypz}). These maximum slope changes are about equal in magnitude (the change is only slightly smaller at the pole compared to the area immediately surrounding it, but not enough to make a difference to the avalanches), so for simplicity we just use the poles themselves as our testing points. Third, the duration of strong tidal effects depends on the eccentricity of the encounter trajectory. We checked the trajectories shown in \fig{traject} and found the duration of strong perturbation, defined for illustration as the period when the force magnitude stays above $90\%$ of the peak value, is tens of minutes long; correspondingly, the responses of the surface material are also transitory and weak.

\begin{center}
\Fig{patterns}.
\end{center}

The three poles p$_x$, p$_y$ and p$_z$ (see \Fig{traject}) were chosen as the test locations for the sample sandpiles since the effective gravity slope angle is near zero at these locations, which enables the sandpiles to hold their initial shapes. \Fig{pxpypz} illustrates the time variation of slope angle at the poles during the encounter. The results derived from all $12$ trajectories in \fig{traject} are shown and to the same scale. The most perturbed pole is p$_y$ on the medium axis, which gains the largest slope angle change for most cases, except in trajectories \{3,4,9,10\}. It is notable that trajectories in the same plane always share similar variation at the three poles, such as \{1,2,5,6\}, \{3,4,9,10\}, and \{7,8,11,12\}. The curves at p$_x$ and p$_y$ show similar doublet shapes since those poles are both located in the rotational plane. The variation at p$_z$ is more complicated due to the fact that the spin pole is not perpendicular to the orbit plane. 

\begin{center}
\Fig{pxpypz}.
\end{center}

\Fig{pxpypz} can be used to estimate the changes in the mechanical environment at these pole locations. Since the slope angle change is the primary mechanism to drive avalanches on the sandpiles, the results derived from different trajectories serve as a framework to determine the magnitude of tidal perturbation at the candidate locations and locations in between, which turns out to be quite small for the Apophis encounter (less than about $1^{\circ}$ in slope for these cases).

\subsection{Tidal Disturbances in 2029}

As analyzed above, catastrophic avalanches solely due to external forces acting directly on the surface particles may never occur on Apophis during the 2029 encounter since the tidal perturbation will be very weak, however it might still have the potential to trigger some local tiny landslides of the surface materials. Generally, the regolith experiences more activity than the constituents deep in the asteroid due to the dynamical environment on the surface, including the microgravity, maximum tidal and centrifugal acceleration, and smaller damping forces from the surroundings due to the smaller confining pressure, therefore we conclude that resurfacing due to external perturbations should occur before wholesale reshaping during the encounter (discussed in \sect{reshpeff}). In the following sections we detail our numerical examination of the local sandpiles for the predicted Apophis encounter scenario to estimate the limit and magnitude of the material responses in 2029. 

One problem that must be confronted in the numerical simulations is that the soft-sphere sandpiles exhibit slow outward spreading due to the accumulation of velocity noise, and this slow spreading may eventually lead to a collapse of the whole sandpile. Unfortunately, our integration time required is long (hours) for a granular system, thus the numerical noise has to be well limited in our method. Two techniques are used in this study. First, as described in \sect{initialcondition}, a rough pallet made up of closely packed spheres in a disk fixed to the ground is used to reduce spreading of the bottom particles. Second, we introduce a critical speed in the code, below which all motions are considered to be noise and are forced to zero. We found a speed threshold $\sim 10^{-10}$ m/s by launching hours-long simulations of a static sandpile for different material properties and finding the corresponding minimum value of the critical speed that limits the numerical spreading. 

Simulations of sandpiles without a flyby were carried out first, serving as a reference for subsequent flyby simulations. Sandpiles in this situation were confirmed to stay equilibrated for a long period, which suggests the avalanches in our numerical experiments (if any) would be attributed entirely to the effects of the tidal encounter. We performed local simulations with the sandpiles generated in \sect{initialcondition} to consider the response of different materials to the Apophis encounter. The local frames at the three poles p$_x$, p$_y$ and p$_z$ (\fig{traject}) were used to position the sample sandpiles, and the flyby simulation data derived from the $12$ fiducial trajectories were employed as the source of external perturbations for the local simulations. There are $108$ possible combinations in total for different materials, locations, and trajectory orientations, covering a wide range of these undetermined parameters. In all the simulations, the start distance of Apophis from the center of the planet was set to $18$ $\mathrm{R_E}$ to ensure the sandpiles were equilibrated fully before it approached the perigee. For our study, we concentrated on the connections between the sandpile's responses and significant variables, \eg spin orientation and sandpile locations, by which we will get a better sense of the surface processes due to a tidal encounter. The sandpiles located at pole p$_y$ for trajectory 2 (\fig{traject}) were examined in detail, which is the case suffering the strongest tidal perturbation (\fig{pxpypz}). \Fig{avalanches} illustrates the disturbances on sandpiles of three different materials for this configuration. Each panel includes a snapshot of the sandpile after the tidal encounter with the displaced particles highlighted and a diagram showing the time variations of the total kinetic energy with planetary distance. Intensive events can be identified by noting the peaks of the kinetic energy curves, and the particles involved in these events are marked with different colors to show the correspondence between the disturbed site and occurrence time. 

\begin{center}
\Fig{avalanches}.
\end{center}

The scale of the disturbances proves to be tiny even for the ``strongest" case in \fig{avalanches}. As illustrated, the displaced particles are few in number and mostly distributed on the surface of the sandpile. We found that the maximum displacement of a given particle is less than $0.8$ times its radius; that is, these disturbances only resulted from the collapse of some local weak structures and the small chain reaction among the surrounding particles. Moreover, the ``gravel" sandpile proves to be unaffected by the tidal encounter, because the static friction is quite large (\tbl{SSDEMparams}) and all particles are locked in a stable configuration, in which case the total kinetic energy of the sandpile remains very small ($\sim 10^{-14}$ J). The ``glass beads" sandpile suffered small but concentrated disturbances that involved very few particles. Accordingly, the displacements of these particles are relatively large. The ``smooth" sandpile presents near-fuild properties with many surface particles experiencing small-amplitude sloshing. The motion eventually damps out and the displacements of the involved particles are tiny (smaller than that of ``glass beads"). 

Based on the detailed analysis of this representative scenario, we performed simulations at other poles and at other orientations, constructing a database to reveal any connections between disturbances and these parameters. However, the results show little association between these two: for the ``gravel" sandpile, no disturbances were detected for any location and any orientation due to the large $\mu_S$; for the ``glass beads" sandpile, some small-scale avalanches can always be triggered, while the occurrence (site and time) of these avalanches seems to be widely distributed and independent of location and orientation; and for the ``smooth" sandpile, surface particles can feel the tidal perturbation and show slight sloshing, but no visible avalanches eventually resulted, because the initial low slope angle in this case imposes a relatively stable structure that is always able to recover from the external perturbations. 

\subsection{Tide-induced Avalanches at Closer Approaches}

To illustrate the effect of stronger perturbations on our model sandpiles, we carried out a few simulations of closer approaches, specifically at $4.0$ and $2.0$ $\mathrm{R_E}$, for the same encounter speed of $8.4$ km/s. These scenarios,  though unlikely to occur in 2029 based on current understanding of Apophis' orbit, are presented here to illustrate the magnitudes of tidal resurfacing effects over a wider range of perturbation strengths. 

Flyby simulations of rubble piles were first carried out to quantify the reshaping effects, using the monodisperse and bidisperse particle rubble-pile models (see \sect{reshpeff}). The bulk densities for both models were set to $\sim 2.4$ g/cc. \tbl{Reshp24} presents the relative net changes of shape factor $\Delta \sigma / \sigma$ for these rubble piles of different materials at different perigee distances. 

\begin{center}
\tbl{Reshp24}.
\end{center}

\tbl{Reshp24} shows results consistent with \sect{reshpeff}, namely that the bimodal rubble pile shows greater fluidity and larger shape changes during the encounter.  To be more specific, the magnitude of reshaping effects at perigee distance $4.0$ $\mathrm{R_E}$ (still larger than the Roche limit of $3.22$ $\mathrm{R_E}$ for a fluid body) remains small ($\sim 10^{-4}$), and that at perigee distance $2.0$ $\mathrm{R_E}$ (smaller than the Roche limit) achieves a significant level of $10^{-3}$--$10^{-1}$. It is notable that the rubble piles did not experience any irreversible distortion even for an approach distance as close as $2.0$ $\mathrm{R_E}$, because the duration of strong tidal effects is relatively short (see \sect{reshpeff}). Anyway, in this section we still adopt the assumption of rigidity to measure the quantities required by the local simulations, which is acceptable since these scenarios are fictitious and only designed to exhibit some massive resurfacing effects. 

For the same reason, in this section we do not present systematic local simulations for different sandpile locations and orientations as done for the real encounter scenario; instead, we simply place the sample sandpiles (see \fig{sandpile}) at the pole of the body long axis p$_x$, and choose trajectory $1$ (see \fig{traject}) as the encounter trajectory to give rise to the maximum tidal effects. \Figtwo{refrsh4f}{refrsh2f} illustrate the responses of the three sandpiles for perigee distance $4.0$ $\mathrm{R_E}$ and $2.0$ $\mathrm{R_E}$, respectively. Each panel includes snapshots of the sandpile before and after the tidal encounter, and the overall shape of the sandpile is traced with a white border for emphasis. Diagrams showing the time variation of the total kinetic energy with planetary distance are also included, in which the intensive avalanches can be identified by noting the peaks of the kinetic energy curves (note the different vertical scales). 

\begin{center}
\Fig{refrsh4f}.
\end{center}

\begin{center}
\Fig{refrsh2f}.
\end{center}

As illustrated in \fig{refrsh4f}, the shape changes of the three sandpiles are still small (but visible) for the encounter at $4.0$ $\mathrm{R_E}$ perigee, and involve many more particles than the encounter at $5.6$ $\mathrm{R_E}$ (\fig{avalanches}). And accordingly, the magnitudes of the total kinetic energy increase consistently for the three sandpiles of different materials, except for the gravel case at $4.0$ $\mathrm{R_E}$, where it appears the frictional ``lock'' established when the pile was first created is disturbed enough to cause a stronger distortion than might otherwise be expected, causing a sharp spike in the kinetic energy plot. Essentially, the ``gravel'' sandpile experienced a more significant collapse than the ``glass bead'' sandpile. As stated in \sect{initialcondition}, the three sandpiles were constructed in a manner that allowed for some inherent randomness, thus their energy state and strength may differ to some degree, and appearently the ``gravel'' sandpile was closer to its failure limit than the other two. This adds an extra element of stochasticity to the results, an aspect to be explored in future work. 

\Fig{refrsh2f} shows the results from the encounter at $2.0$ $\mathrm{R_E}$ perigee, for which the encounter trajectory partly entered the Roche sphere ($3.22$ $\mathrm{R_E}$). In this case, the shapes of the three sandpiles are highly distorted during the encounter, with the involved particles slumped towards the direction where the planet is receding. The corresponding changes in total kinetic energy become extremely large when the massive avalanches occur, especially for the ``smooth'' particles. 

The results of this section suggest that a $4.0$ $\mathrm{R_E}$ encounter may alter the regolith on Apophis' surface slightly, and a $2.0$ $\mathrm{R_E}$ encounter may produce a strong resurfacing effect (of course, we would expect considerable global distortion as well, if the asteroid is a rubble pile). The shape changes of the sandpiles depend on the orientation of the encounter trajectory; \ie particles can be dragged away along the direction in which the planet recedes. 

\section{DISCUSSION} \label{s:discuss}

The argument about whether a terrestrial encounter can reset the regolith of NEAs has been discussed for a while (\eg Binzel \etal\ 2010, Nesvorn\'y \etal\ 2010). Important evidence is provided by measurements of the spectral properties, which suggest that asteroids with orbits conducive to tidal encounters with planets show the freshest surfaces, while quantitative evaluation of this mechanism is still rare and rough. The two-stage approach presented in this paper enables the most detailed simulations to date of regolith migration due to tidal effects, which we have applied to make a numerical prediction for the surface effects during the 2029 approach of (99942) Apophis. 

We confirm that the shape modification of Apophis due to this encounter will likely be negligibly small based on the results of systematic simulations parameterized by bulk density, internal structure, and material properties. The analysis of the global mechanical environment over the surface of Apophis during the 2029 encounter reveals that the tidal perturbation is even too small to result in any large-scale avalanches of the regolith materials, based on a plausible range of material parameters and provided that external forces acting directly on the surface dominate any surface effects due to seismic activity emanating from other regions of the body. Future work will explore these second-order surface effects and their relevance to this encounter, and to cases of tidal resurfacing of small bodies in general. It is notable that our approach is capable of capturing slight changes in the regolith (modeled as sandpiles); thus, through numerical simulation, we find that this weak tidal encounter does trigger some local disturbances for appropriate material properties. 

These possible disturbances, though local and tiny, are essentially related to the nature of the sandpile, which is actually a bunch of rocks and powders in a jammed state. The sandpile is formed by a competition between the constituents to fall under gravity and eventually reach an equilibrium with mutual support. It can hold a stable shape under a constant environment force at low sandpile densities (compared with close packing), which is primarily due to interior collective structures, called bridges, that are formed at the same time as the sandpile and are distributed non-uniformly throughout the sandpile (Mehta 2007). The soft-sphere method provides a detailed approximation to a real sandpile in terms of the granular shape and contact mechanics, which can well reproduce the bridges that dominate the structure of the sandpile. Accordingly, we can describe the nuanced responses due to the changes in $(F, \theta, \alpha)$---see \sect{GCME} and \fig{ACCS}. Changing the slope angle $\theta$ is the most efficient way to break the equilibrium of substructures in the sandpile, while changing the magnitude $F$ and deflection angle $\alpha$ may also play a role in the causes of avalanches. Changes in $F$ can readjust the interparticle overlaps and cause collapse of some bridges, which is a principal mechanism of compaction by filling the voids during small structural modifications (Mehta \etal\ 2004). This dependence was recently demonstrated in experiments of measuring the angle of repose under reduced gravity (Kleinhans \etal\ 2011). Similarly, the weak links in bridges of the sandpile may also be disturbed and broken during the sweeping of $\alpha$ in the plane. In addition, Mehta \etal\ (1991) pointed out that even slight vibration caused by the environment force may result in collapse of long-lived bridges, which is also responsible for the sandpile landslides during the encounter. 

Although the magnitudes of the avalanches are different for the three materials we tested, they are nonetheless all very small. We note the occurrence of these local avalanches shows little dependence on the orientation/location of the sandpile, because all the perturbations are small in magnitude and the sandpile behavior, these small-scale avalanches, actually depends on the presence of weak bridges inside the sandpile, which in turn could be sensitive to the changes of the environmental force's magnitude, deflection angle, and slope angle in a somewhat random way.

In any case, this numerical study shows that tidal resurfacing may not be particularly effective at moderate encounter distances. We predict that overall resurfacing of Apophis, regolith will not occur if the only source of disturbance is external perturbations. Mini-landslides on the surface may still be observed by a visiting spacecraft with sufficiently sensitive monitoring equipment. This provides a great motivation for $in$ $situ$ exploration of Apophis in 2029 (Michel \etal\ 2012).

\section{CONCLUSIONS} \label{s:concl}

This paper provides a numerically derived prediction for the surface effects on (99942) Apophis during its 2029 approach, which is likely to be one of the most significant asteroid encounter events in the near future.  A two-stage scheme was developed based on the soft-sphere code implementation in \code{pkdgrav} to mimic both a rubble pile's (rigid and flexible) responses to a planetary flyby and a sandpile's responses to all forms of perturbations induced by the encounter. The flyby simulations with the rubble pile indicate that reshaping effects due to the tidal force on Apophis in 2029 will be negligibly small for bulk densities in the expected range ($2$--$3$ $\mathrm{g/cm^3}$). The resultant environmental force felt by the sandpile on the asteroid surface was approximated with a uniform analytical expression, which led to an estimate of the changes in the global mechanical environment. Three typical patterns of perturbation were presented based on the asteroid body and spin orientation at perigee, showing a general dependence of the magnitude of tidal perturbation on the orientation of the trajectory. Twelve fiducial trajectories were used to calculate the magnitude of the tidal perturbation at three poles of the tri-axial ellipsoid model, indicating that the strongest tidal perturbation appears where the local slope is originally steep and that the duration of the strong perturbation is short compared with the whole process. The tidal perturbation on surface materials is confirmed to be quite weak for the 2029 encounter, therefore large-scale avalanches may never occur. However, we showed this weak perturbation does trigger some local tiny landslides on the sample sandpiles, though the involved particles are few in number and are distributed on the surface of the sandpile. These small-scale avalanches result from the breaking of weak substructures by slight external perturbations, therefore the occurrence of these local avalanches is widely distributed and presents little dependence on the encounter parameters. The simulations of closer approaches show that an encounter at $2.0$ $\mathrm{R_E}$ is capable of triggering some massive avalanches of the sandpiles, \ie to alter the regolith on Apophis' surface significantly (although the entire body would also undergo significant shape change in this case).

Further research will be performed to generalize our work over a wide range of possible asteroid-planet encounter conditions. And as stated above, we will also investigate whether even slight internal perturbations in the asteroid during tidal encounters may contribute to noticeable surface motions. 

\newpage
\section*{Acknowledgments}

This material is based on work supported by the US National Aeronautics and Space Administration under Grant Nos.\ NNX08AM39G, NNX10AQ01G, and NNX12AG29G issued through the Office of Space Science and by the National Science Foundation under Grant No.\ AST1009579. Some simulations in this work were performed on the ``yorp" cluster administered by the Center for Theory and Computation in the Department of Astronomy at the University of Maryland. Patrick Michel acknowledges support from the French space agency CNES. Ray tracing for Figs.1, 2, 4, 5 and 12 was performed using POV-Ray (http://www.povray.org/).

\newpage
\section*{References}

\begin{description}

\item \paper {Benz, W., Asphaug, E.} 1999 {Catastrophic disruptions revisited} Icarus 142 5--20

\item \paper {Binzel, R.P., Rivkin, A.S., Thomas, C.A., Vernazza, P., Burbine, T.H., DeMeo, F.E., Bus, S.J., Tokunaga, A.T., Birlan, M.} 2009 {Spectral properties and composition of potentially hazardous Asteroid (99942) Apophis} Icarus 200 480--485

\item \paper {Binzel, R.P., Morbidelli, A., Merouane, S., DeMeo, F.E., Birlan, M., Vernazza, P., Thomas, C.A., Rivkin, A.S., Bus, S.J., Tokunaga, A.T.} 2010 {Earth encounters as the origin of fresh surfaces on near-Earth asteroids} Nature 463 331--334

\item \book {Brown, R.L., Richards, J.C.} 1966 {Principles of Powder Mechanics} {Oxford: Pergamon}

\item \aIII {Clark, B.E., Hapke, B., Pieters, C., Britt, D.} {Asteroid space weathering and regolith evolution} {485--500}

\item \paper {Delbo', M., Cellino, A., Tedesco, E.F.} 2007 {Albedo and size determination of potentially hazardous asteroids: (99942) Apophis} Icarus 188 266--269

\item \paper {DeMeo, F.E., Binzel, R.P., Lockhart, M.} 2013 {Mars encounters cause fresh surfaces on some near-Earth asteroids} Icarus 227 112--122

\item \paper {Fujiwara, A. Kawaguchi, J., Yeomans, D.K., Abe, M., Mukai, T., Okada, T., Saito, J., Yano, H., Yoshikawa, M., Scheeres, D.J., Barnouin-Jha, O., Cheng, A.F., Demura, H., Gaskell, R.W., Hirata, N., Ikeda, H., Kominato, T., Miyamoto, H., Nakamura, A.M., Nakamura, R., Sasaki, S., Uesugi, K.} 2006 {The rubble-pile asteroid Itokawa as observed by Hayabusa} Science 312 1330--1334

\item \paper {Giorgini, J.D., Benner, L.A.M., Ostro, S.J., Nolan, M.C., Busch, M.W.} 2008 {Predicting the Earth encounters of (99942) Apophis} Icarus 193 1--19

\item \paper {Holsapple, A., Michel, P.} 2006 {Tidal disruptions: A continuum theory for solid bodies} Icarus 183 331--348

\item \paper {Hu, W., Scheeres, D.J.} 2004 {Numerical determination of stability regions for orbital motion in uniformly rotating second degree and order gravity fields} {Planetary and Space Science} 52 685--692

\item \paper {Jutzi, M., Michel, P., Benz, W., Richardson, D.C.} 2010 {Fragment properties at the catastrophic disruption threshold: The effect of the parent body's internal structure} Icarus 207 54--65

\item \paper {Kleinhans, M.G., Markies, H., de Vet, S.J., in 't Veld, A.C., Postema, F.N.} 2011 {Static and dynamic angles of repose in loose granular materials under reduced gravity} {Journal of Geophysical Research} 116 E11004

\item \paperL {Nesvorn\'y, D., Bottke Jr., W.F., Vokrouhlick\'y, D., Chapman, C.R., Rafkin, S.} 2010 {Do planetary encounters reset surfaces of near Earth asteroids?} Icarus 209 510--519

\item \book {Mehta, A.} 2007 {Granular Physics} {Cambridge University Press}

\item \paper {Mehta, A., Barker, G.C.} 1991 {Vibrated powders: A microscopic approach} {Physics Review Letter}  67 394--397

\item \paperS {Mehta, A., Barker, G.C., Luck, J.M.} 2004 {Cooperativity in sandpiles: Statistics of bridge geometries} {Journal of Statistical Mechanics: Theory and Experiment}  P10014

\item \paper {Michel, P., Benz, W., Tanga, P., Richardson, D.C.} 2001 {Collisions and gravitational reaccumulation: Forming asteroid families and satellites} Science 294 1696--1700

\item \paperC {Michel, P., Prado, J.Y., Barucci, M.A., Bousquet, P., Chiavassa, F., Groussin, O., Herique, A., Hestroffer, D., Hinglais, E., Lopes, L., Martin, T., Mimoun, D., R\'eme, H., Thuillot, W.} 2012 {Probing the interior of asteroid Apophis: A unique opportunity in 2029} {Special Session 7 of the XXVIII IAU General Assembly, 29--31 August, 2012, Beijing (China), Abstract Book}

\item \paper {Richardson, D.C., Bottke, Jr., W.F., Love, S.G.} 1998 {Tidal distortion and disruption of Earth-crossing asteroids} Icarus 134 47--76

\item \paper {Richardson, D.C., Quinn, T., Stadel, J., Lake, G.} 2000 {Direct large-scale $N$-body simulations of planetesimal dynamics} Icarus 143 45--59

\item \aIII {Richardson, D.C., Leinhardt, Z.M., Melosh, H.J., Bottke Jr., W.F., Asphaug, E.} {Gravitational aggregates: Evidence and evolution} {501--515}

\item \paper {Richardson, D.C., Michel, P., Walsh, K.J., Flynn, K.W.} 2009 {Numerical simulations of asteroids modeled as gravitational aggregates} {Planetary and Space Science} 57 183--192

\item \paper {Richardson, D.C., Walsh, K.J., Murdoch, N., Michel, P.} 2011 {Numerical simulations of granular dynamics: I. Hard-sphere discrete element method and tests} Icarus 212 427--437

\item \paper {Richardson, D.C., Blum, J., Weinhart, T., Schwartz, S.R., Michel, P., Walsh, K.J.} 2012 {Numerical simulations of landslides calibrated against laboratory experiments for application to asteroid surface processes} {American Astronomical Society, DPS meeting}  44 105.06

\item \paper {Scheeres, D.J., Benner, L.A.M., Ostro, S.J., Rossi, A., Marzari, F., Washabaugh, P.} 2005 {Abrupt alteration of Asteroid 2004 MN4's spin state during its 2029 Earth flyby} Icarus 178 281--283

\item \paper {Schwartz, S.R., Richardson, D.C., Michel, P.} 2012 {An implementation of the soft-sphere discrete element method in a high-performance parallel gravity tree-code} {Granular Matter} 14 363--380

\item \paper {Schwartz, S.R., Michel, P., Richardson, D.C.} 2013 {Numerically simulating impact disruptions of cohesive glass bead agglomerates using the soft-sphere discrete element method} Icarus 226 67--76 

\item \inpress {Schwartz, S.R., Michel, P., Richardson, D.C., Yano, H.} 2014 {Low-speed impact simulations into regolith in support of asteroid sampling mechanism design I.: Comparison with 1-g Experiments} {Planet. Space Sci.}

\item \thesis {Stadel, J.} 2001 {Cosmological $N$-body simulations and their analysis} {University of Washington} 141

\item \paper {Tanga, P., Comito, C., Paolicchi, P., Hestroffer, D., Cellino, A., Dell’Oro, A., Richardson, D. C., Walsh, K. J., Delbo', M.} 2009 {Rubble-pile reshaping reproduces overall asteroid shapes} {The Astrophysical Journal} 706 197--202

\item \paper {Tholen, D.J., Micheli, M., Elliott, G.T.} 2013 {Improved astrometry of (99942) Apophis} {Acta Astronautica} 90 56--71

\item \paper {Veverka, J., and 32 co-authors} 2000 {NEAR at Eros: Imaging and spectral results} Science 289 2088--2097

\item \paper {Walsh, K. J., Richardson, D.C., Michel, P.} 2012 {Spin-up of rubble-pile asteroids: Disruption, satellite formation, and equilibrium shapes} Icarus 220 514--529

\item \paper {Wlodarczyk, I.} 2013 {The potentially dangerous asteroid (99942) Apophis} {Monthly Notices of the Royal Astronomical Society} 434(4) 3055--3060

\end{description}

%
%
\newpage
\renewcommand{\baselinestretch}{1} 

\begin{table}[h]
  \centering
  \caption{Soft-sphere parameter sets used to represent three different material
    properties of sandpiles in the simulations.  Here $\mu_s$ is the
    coefficient of static friction ($= \tan^{-1}\phi$, where $\phi$ is
    the material friction angle), $\mu_r$ is the coefficient of
    rolling friction, $\varepsilon_n$ is the normal coefficient of
    restitution (1 = elastic, 0 = plastic), $\varepsilon_t$ is the
    tangential coefficient of restitution (1 = no sliding friction), $k_n$ is the normal spring constant ($\mathrm{kg/s^2}$), and $k_t$ is the tangential spring constant (also $\mathrm{kg/s^2}$). }
  \label{t:SSDEMparams}
  \vspace{0.5in}
  \begin{tabular}{c|ccc}
    Parameter & Gravel & Glass Beads & Smooth\\
    \hline
    $\mu_s$ & 1.31 & 0.43 & 0.0 \\
    $\mu_r$ & 3.0 & 0.1 & 0.0 \\
    $\varepsilon_n$ & 0.55 & 0.95 & 0.95 \\
    $\varepsilon_t$ & 0.55 & 1.0 & 1.0 \\
    $k_n$ & 83.3 & 83.3 & 83.3 \\
    $k_t$ & 23.8 & 23.8 & 23.8 \\
    \hline
    \hline
  \end{tabular}
\end{table}

\newpage
\renewcommand{\baselinestretch}{1} 

\begin{table}[h]
  \centering
  \caption{Relative net change of shape factor $\Delta \sigma / \sigma$ for the flexible rubble-pile model of monodisperse and bidispersed particles, parameterized by the bulk density and material property. The results of bidisperse smooth particles are omitted since the rubble pile in this case cannot hold the overall shape of a 1.4:1.0:0.8 tri-axial ellipsoid at Apophis' spin rate. The estimates of Apophis' Roche limit (second column) corresponding to the indicated bulk density are calculated assuming a fluid body in a circular orbit around Earth. For reference, the Apophis encounter distance in 2029 will be about 5.6 $\mathrm{R_E}$. }
  \label{t:FullScaleChange}
  \vspace{0.5in}
  \begin{tabular}{c|c|c|ccc}
  Density ($\mathrm{g/cm^3}$) & Roche Limit ($\mathrm{R_E}$) & Dispersion & Gravel& Glass Beads & Smooth \\
  \hline
  2.4 & 3.22 & Unimodal & $6.62 \times 10^{-5}$ & $7.76 \times 10^{-5}$ & $3.26 \times 10^{-5}$ \\
  & & Bimodal & $8.36 \times 10^{-4}$& $6.00 \times 10^{-4}$ & --- \\
  \hline
  2.0 & 3.42 & Unimodal & $5.19 \times 10^{-5}$ & $8.84 \times 10^{-5}$ & $8.67 \times 10^{-5}$ \\
  & & Bimodal & $9.67 \times 10^{-4}$ & $6.06 \times 10^{-4}$ & --- \\
  \hline
  1.5 & 3.77 & Unimodal & $2.75 \times 10^{-5}$ & $9.54 \times 10^{-5}$ & $3.60 \times 10^{-5}$ \\
  & & Bimodal & $4.73 \times 10^{-4}$ & $3.51 \times 10^{-4}$ & --- \\
  \hline
  1.0 & 4.31 & Unimodal & $1.88 \times 10^{-5}$ & $4.95 \times 10^{-5}$ & $4.68 \times 10^{-5}$ \\
  & & Bimodal & $9.07 \times 10^{-4}$ & $5.57 \times 10^{-4}$ & --- \\
  \hline
  0.5 & 5.43 & Unimodal & $1.99 \times 10^{-5}$ & $1.45 \times 10^{-5}$ & $2.16 \times 10^{-5}$ \\
  & & Bimodal & $2.82 \times 10^{-3}$ & $2.31 \times 10^{-4}$ & --- \\
  \hline
  0.1 & 9.29 & Unimodal & $2.37 \times 10^{-2}$ & $3.12 \times 10^{-3}$ & $2.47 \times 10^{-2}$ \\
  & & Bimodal & $6.54 \times 10^{-2}$ & $1.31 \times 10^{-1}$ & --- \\
  \hline
  \hline
  \end{tabular}
\end{table}

\newpage
\renewcommand{\baselinestretch}{1} 

\begin{table}[h]
  \centering
  \caption{Relative net change of shape factor $\Delta \sigma / \sigma$ for the rubble piles of monodisperse and bidisperse particles, parameterized by the perigee distance and material property. As before, the results of bidisperse smooth particles are omitted since they cannot maintain Apophis' tri-axial ellipsoidal shape. }
  \label{t:Reshp24}
  \vspace{0.5in}
  \begin{tabular}{c|c|ccc}
  Perigee ($\mathrm{R_E}$) & Dispersion & Gravel & Glass Beads & Smooth \\
  \hline
  2.0  & Unimodal & $1.45 \times 10^{-3}$ & $1.42 \times 10^{-3}$ & $5.62 \times 10^{-1}$ \\
  & Bimodal & $4.48 \times 10^{-2}$ & $7.95 \times 10^{-2}$ & ---\\
  \hline
  4.0  & Unimodal & $1.32 \times 10^{-4}$ & $7.18 \times 10^{-4}$ & $5.88 \times 10^{-4}$ \\
  & Bimodal & $9.37 \times 10^{-4}$ & $8.09 \times 10^{-4}$ & ---\\
  \hline
  \hline
  \end{tabular}
\end{table}

%
%
\newpage
\renewcommand{\baselinestretch}{2} 

\section*{Figure Captions}

\begin{figure}[ht!]
\centering
\subfigure[Sandpile with ``gravel" parameter set.]{
   \label{fig:subfig1}
   \includegraphics[width=0.7\textwidth] {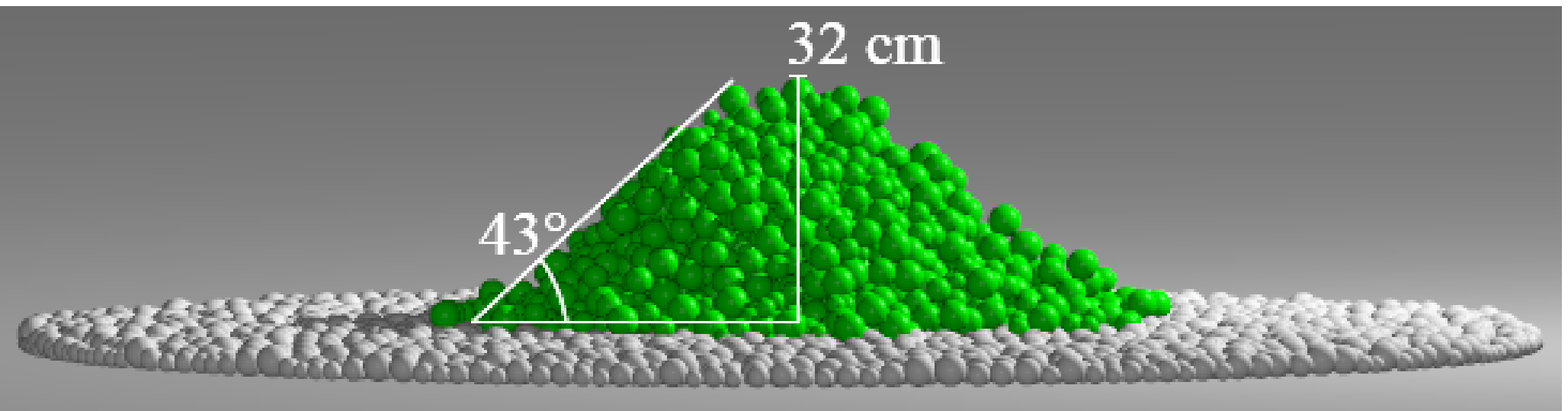}
 }
 \subfigure[Sandpile with ``glass beads" parameter set.]{
   \label{fig:subfig2}
   \includegraphics[width=0.7\textwidth] {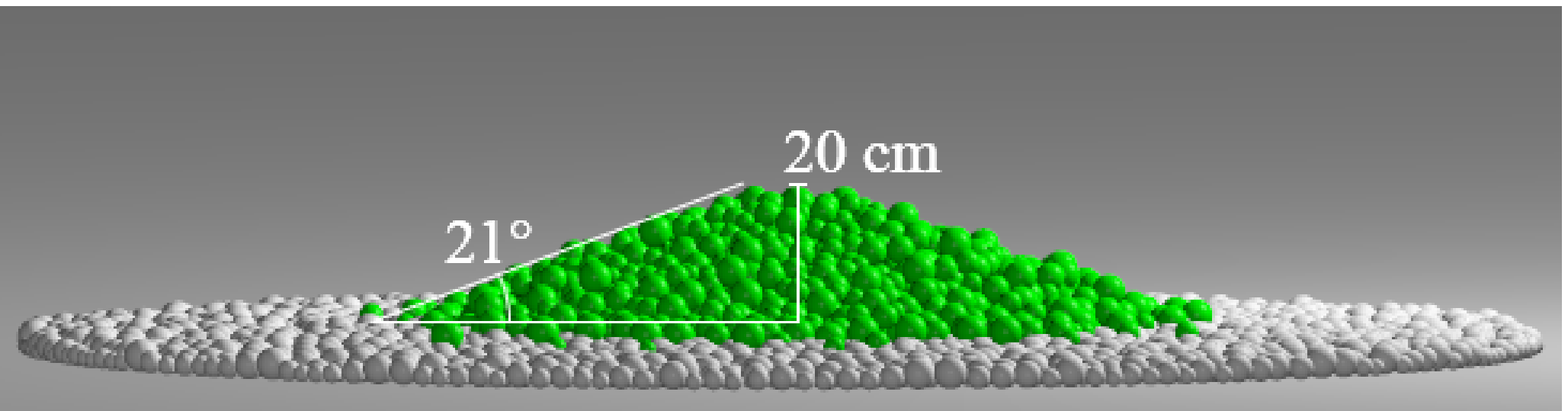} 
 }
 \subfigure[Sandpile with ``smooth" parameter set.]{
   \label{fig:subfig3}
   \includegraphics[width=0.7\textwidth] {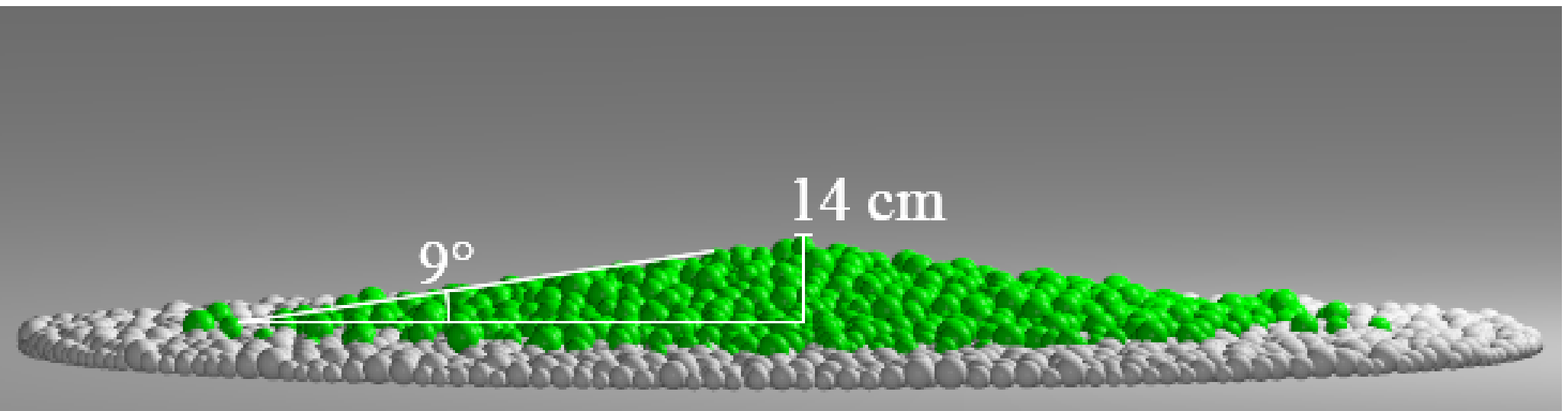}  
 }
\caption{}
\label{f:sandpile}
\end{figure}

\begin{figure}[ht!]
\vspace{0.1in} 
\centering
\subfigure[Experiment: Beginning]{
   \label{fig:subfig1}
   \includegraphics[width=0.3\textwidth] {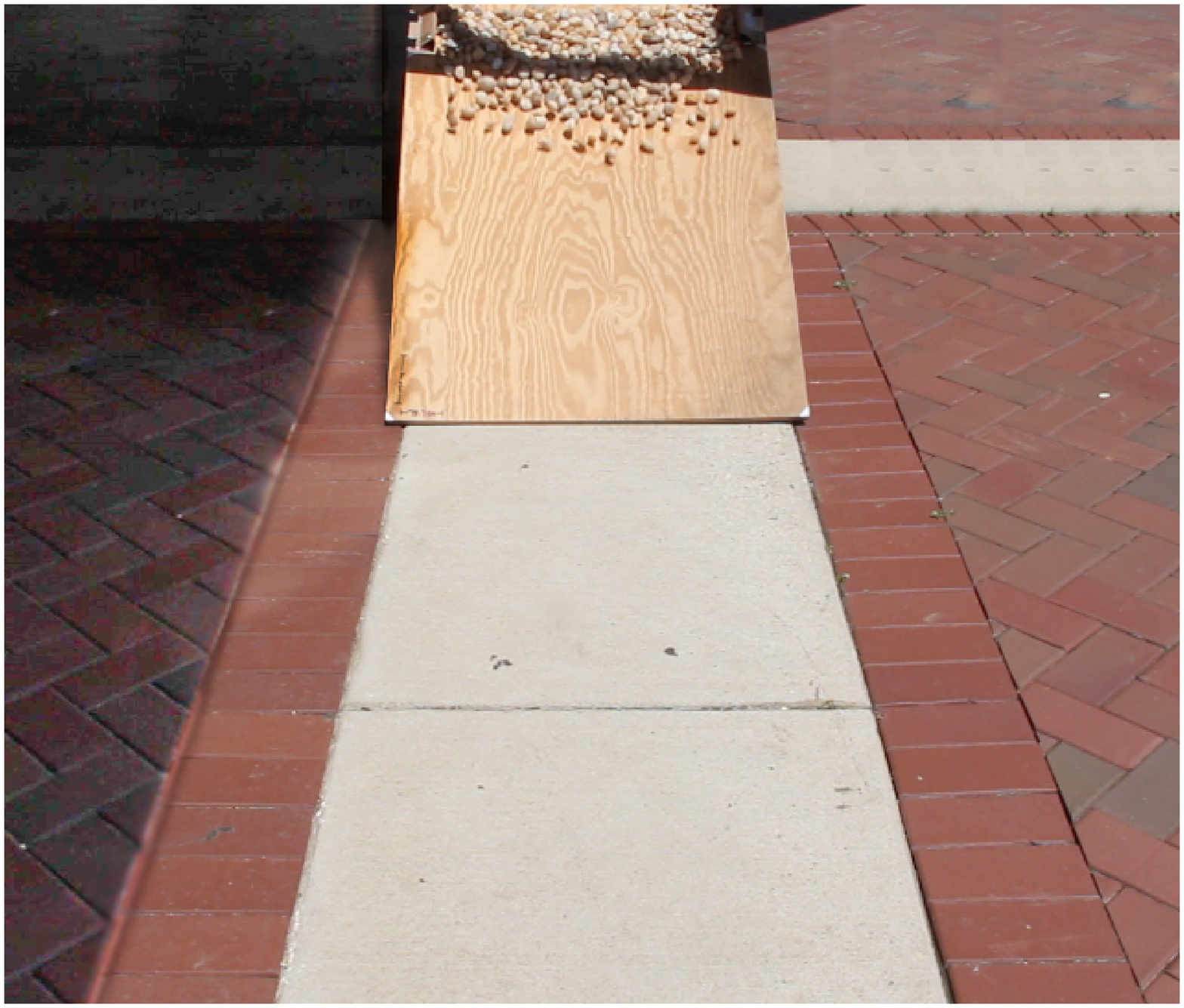}
 }
\subfigure[Experiment: Middle]{
   \label{fig:subfig5}
   \includegraphics[width=0.3\textwidth] {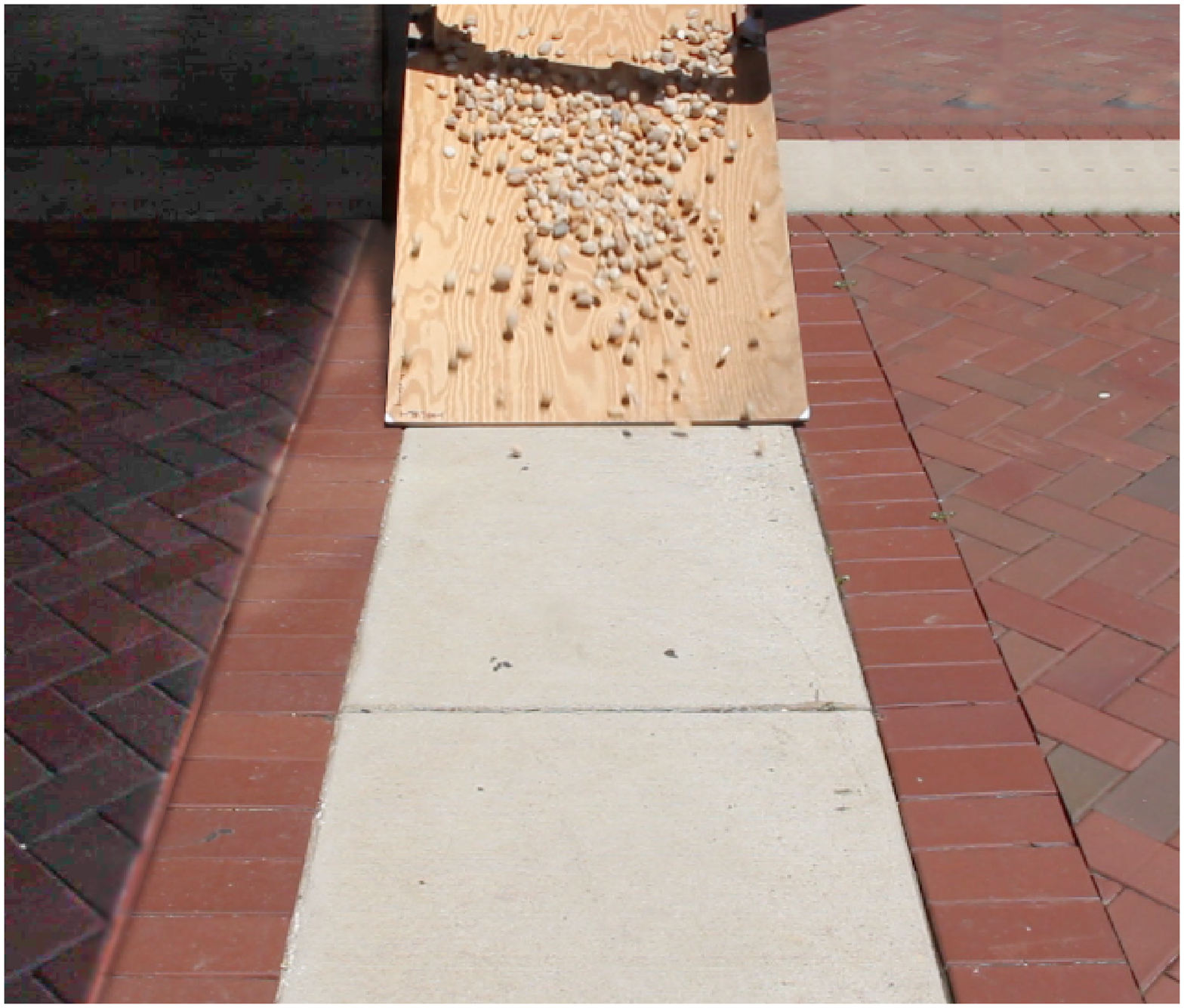}
 }
\subfigure[Experiment: End]{
   \label{fig:subfig3}
   \includegraphics[width=0.3\textwidth] {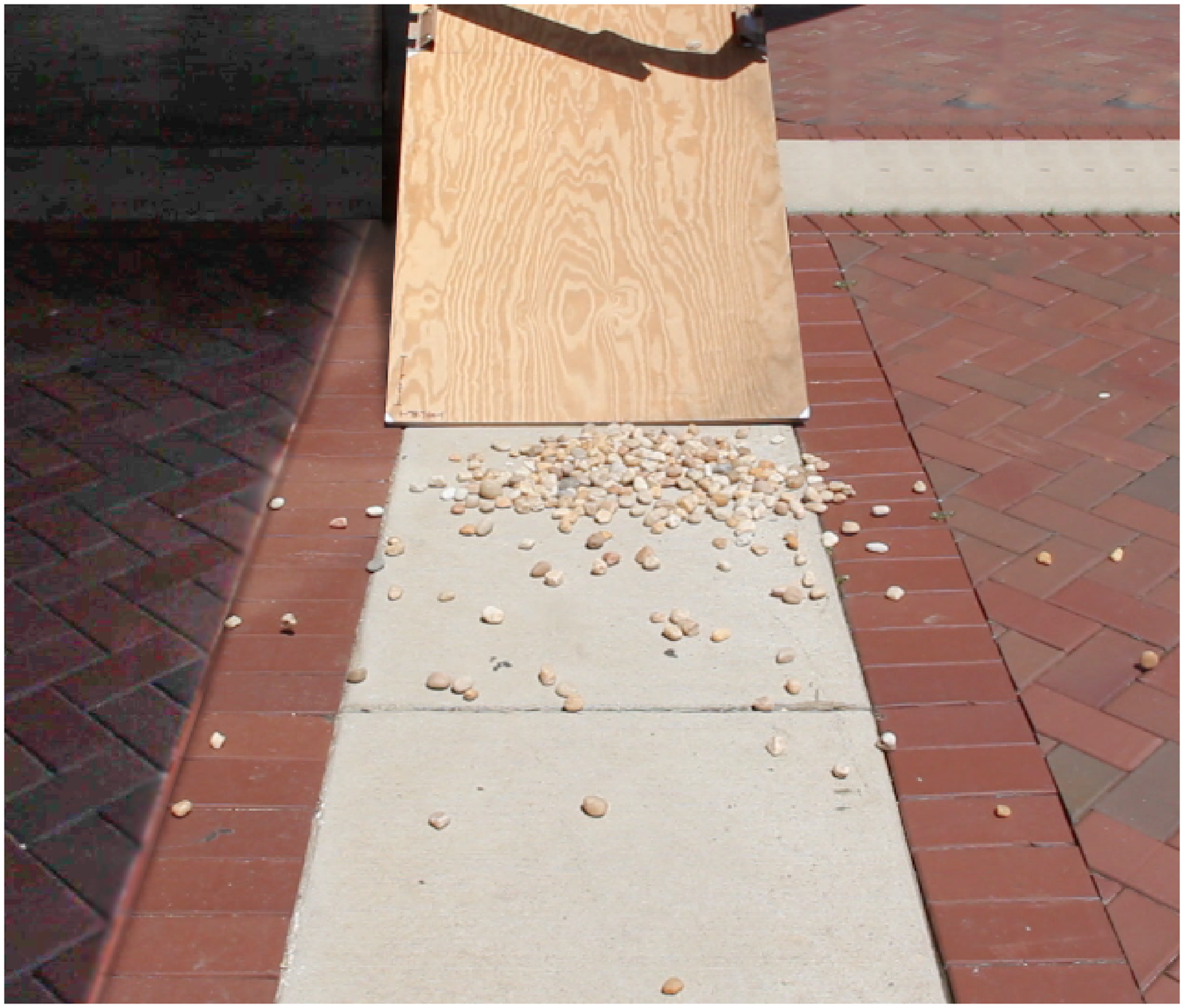}
 }
\subfigure[Simulation: Beginning]{
   \label{fig:subfig2}
   \includegraphics[width=0.3\textwidth] {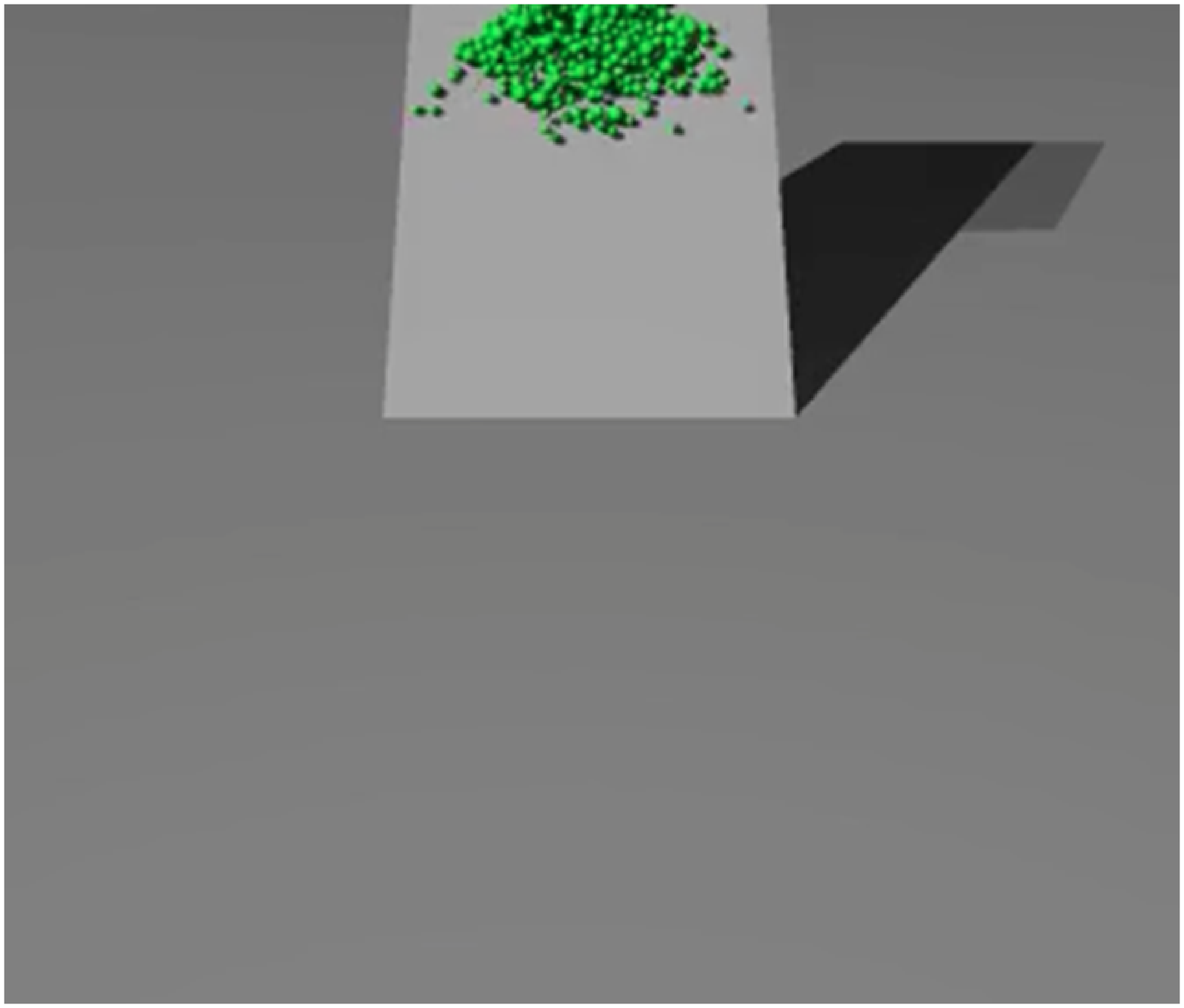} 
}
\subfigure[Simulation: Middle]{
   \label{fig:subfig4}
   \includegraphics[width=0.3\textwidth] {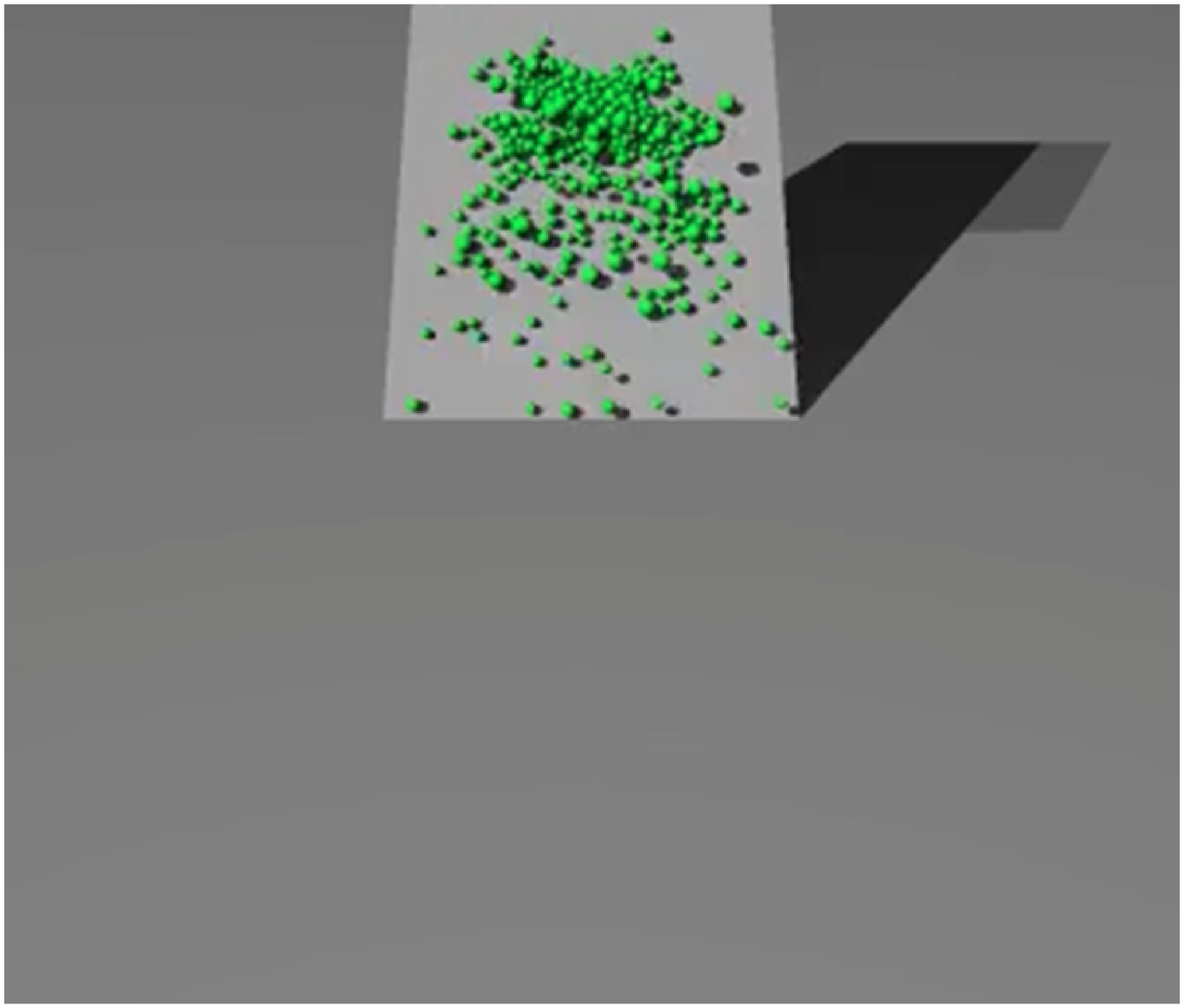} 
}
\subfigure[Simulation: End]{
   \label{fig:subfig6}
   \includegraphics[width=0.3\textwidth] {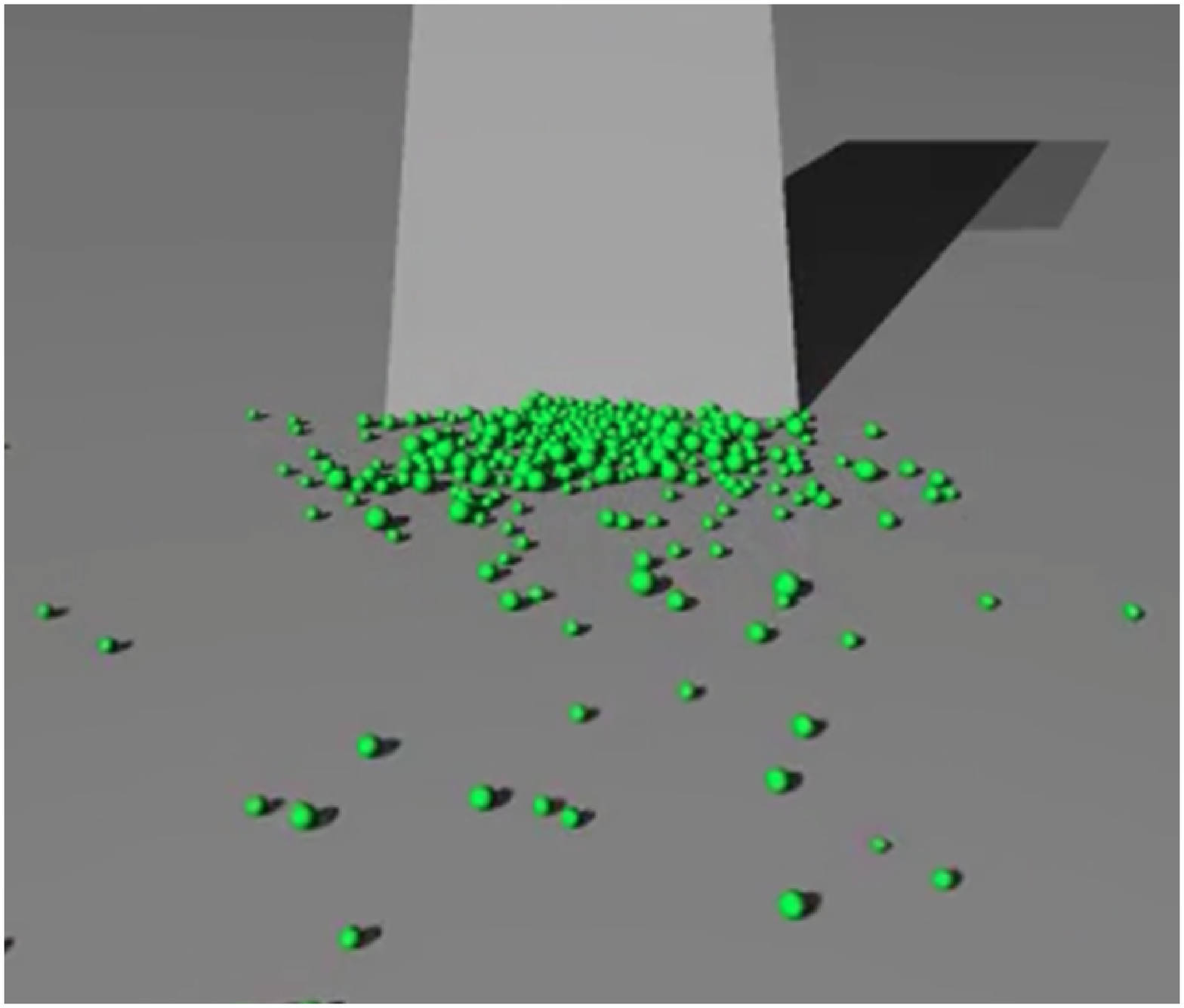} 
 }
\caption{}
\label{f:rockslide}
\end{figure}

\begin{figure}[h]
\centering
\scalebox{0.6}
{\includegraphics{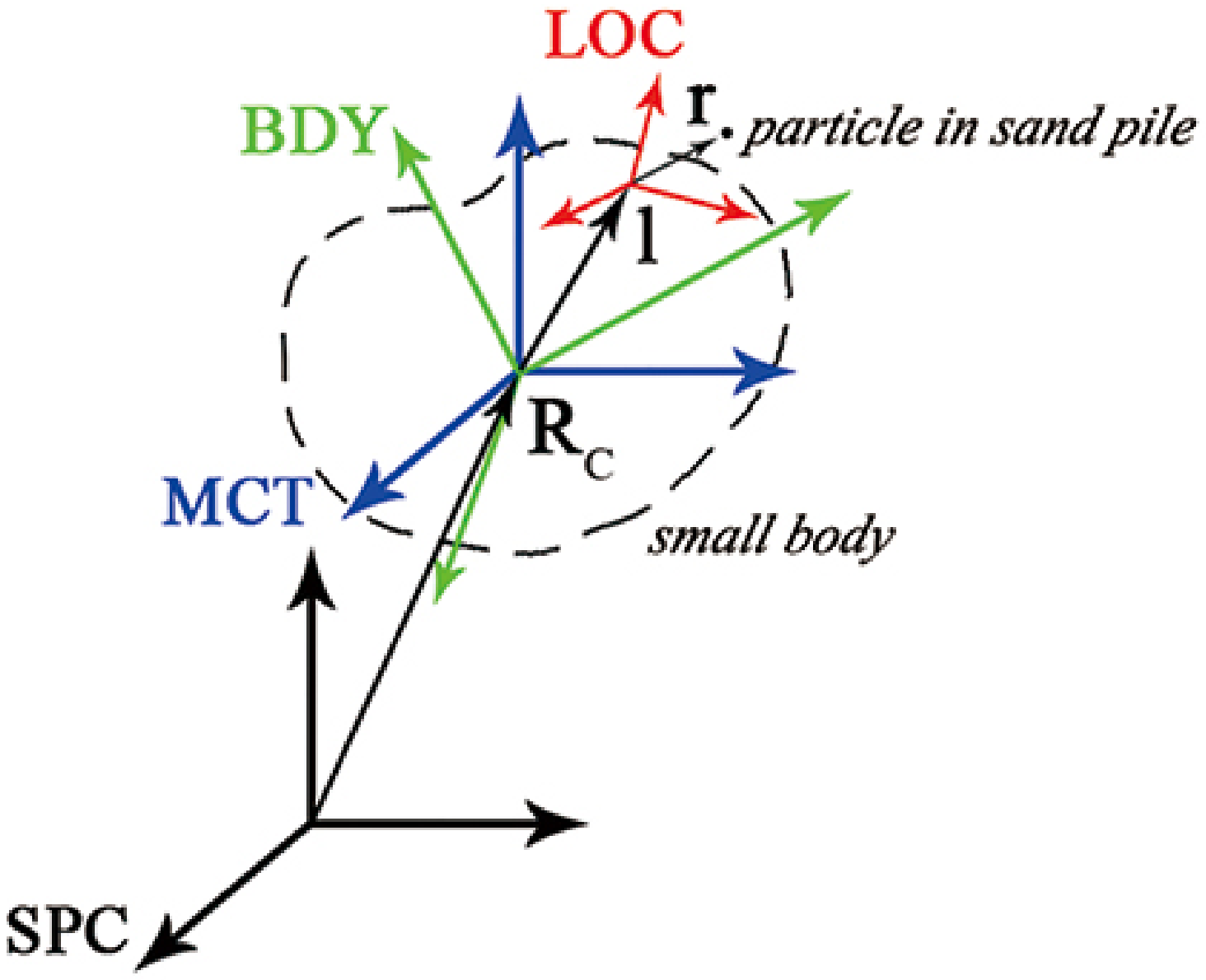}}
\caption{}
\label{f:frames}
\end{figure}

\begin{figure}[h]
\centering
\scalebox{0.45}
{\includegraphics{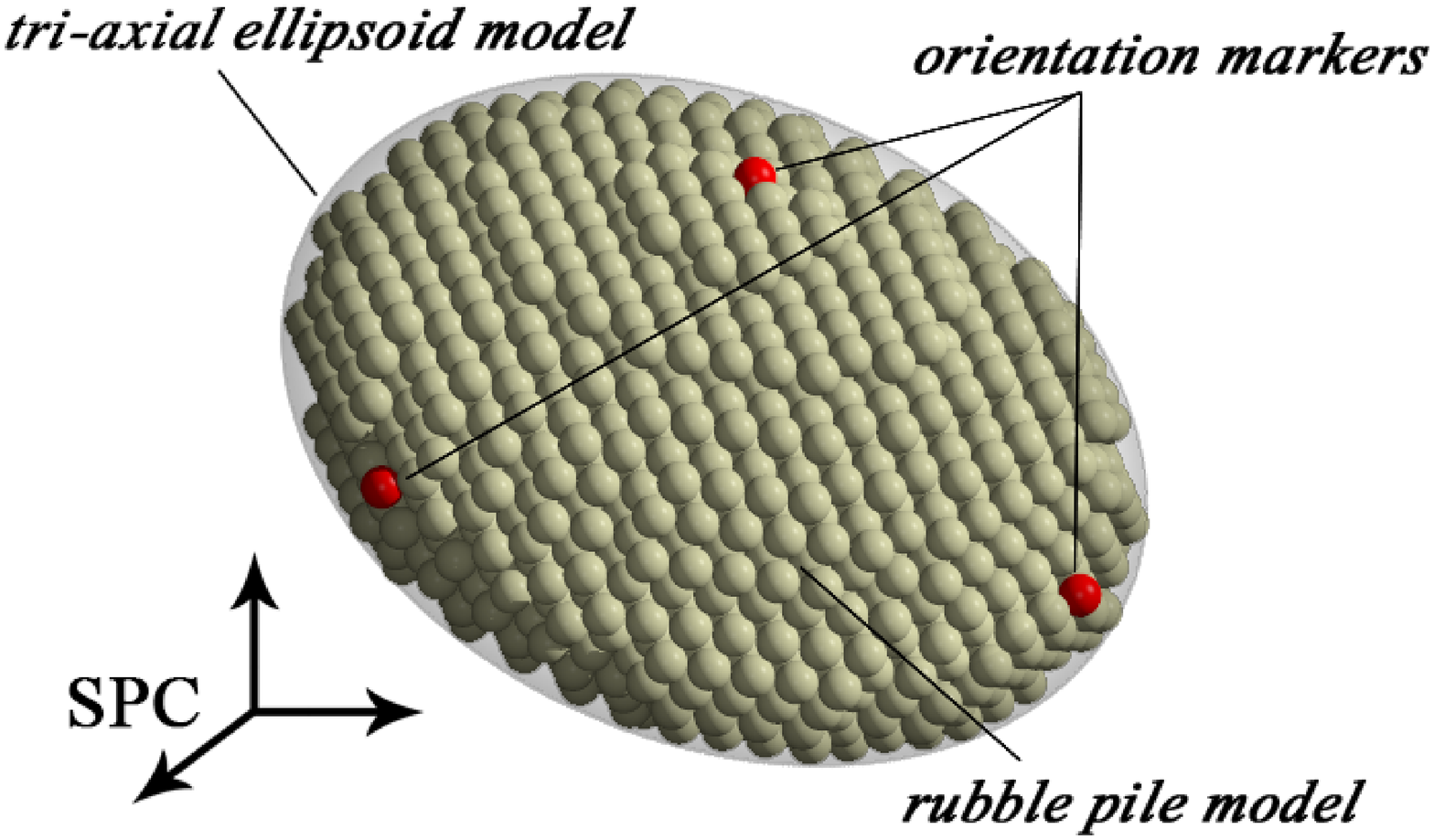}}
\caption{}
\label{f:rubblepile}
\end{figure}

\begin{figure}[h]
\centering
\scalebox{0.4}
{\includegraphics{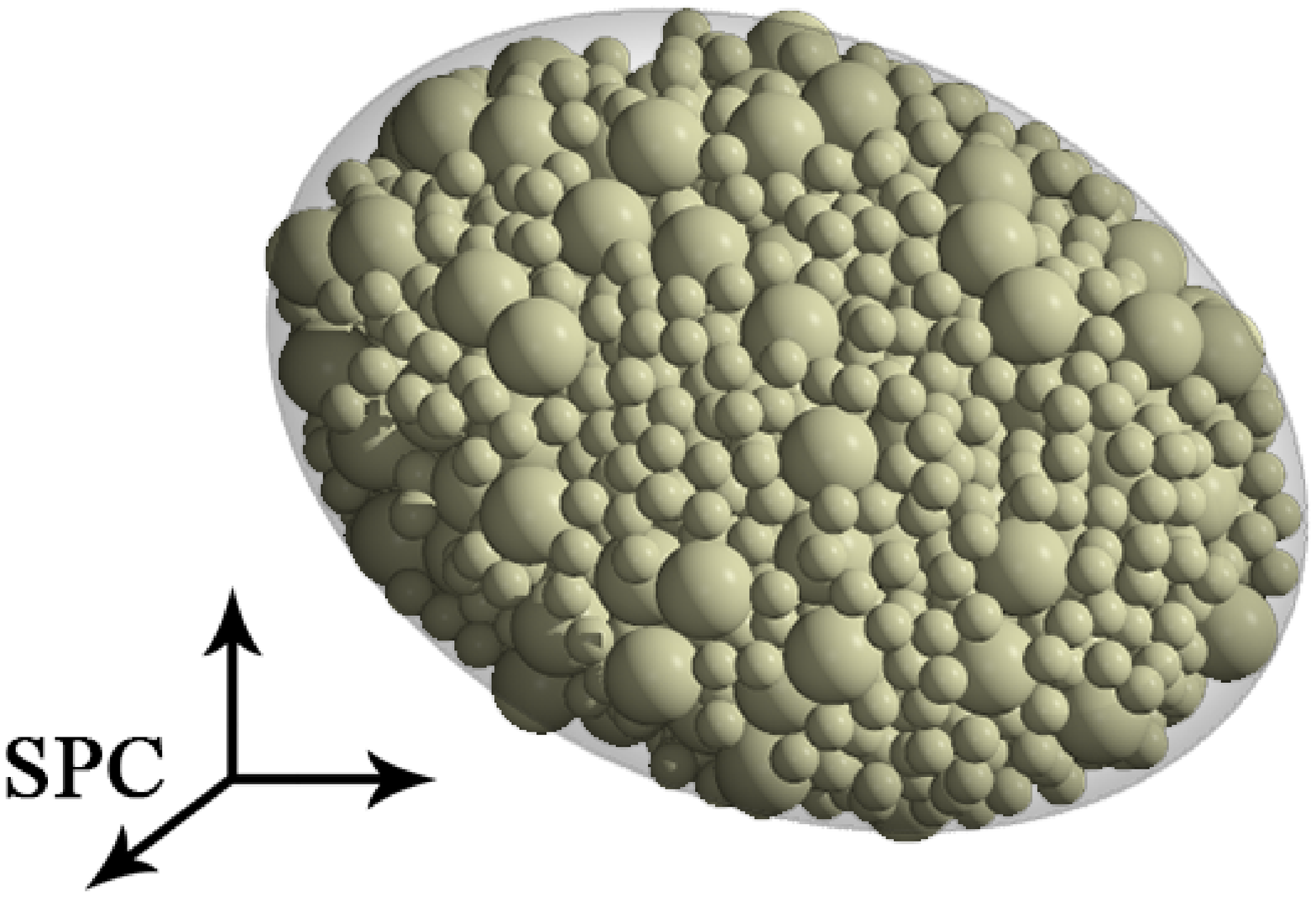}}
\caption{}
\label{f:newpile}
\end{figure}

\begin{figure}[ht!]
\centering
\subfigure[Rubble piles with ``gravel" parameter set.]{
   \label{fig:subfig1}
   \includegraphics[width=0.45\textwidth] {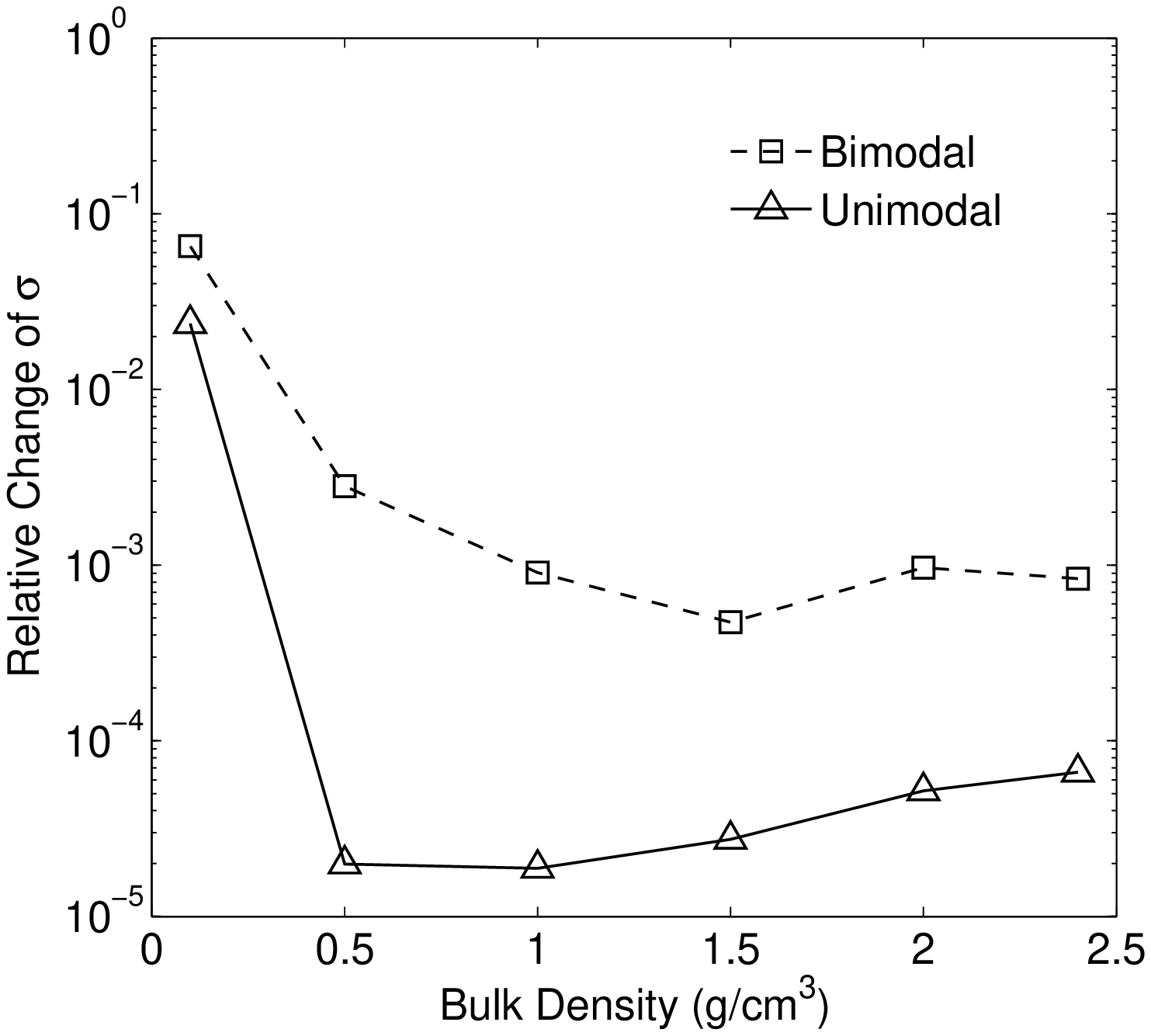}
 }
 \subfigure[Rubble piles with ``glass beads" parameter set.]{
   \label{fig:subfig2}
   \includegraphics[width=0.45\textwidth] {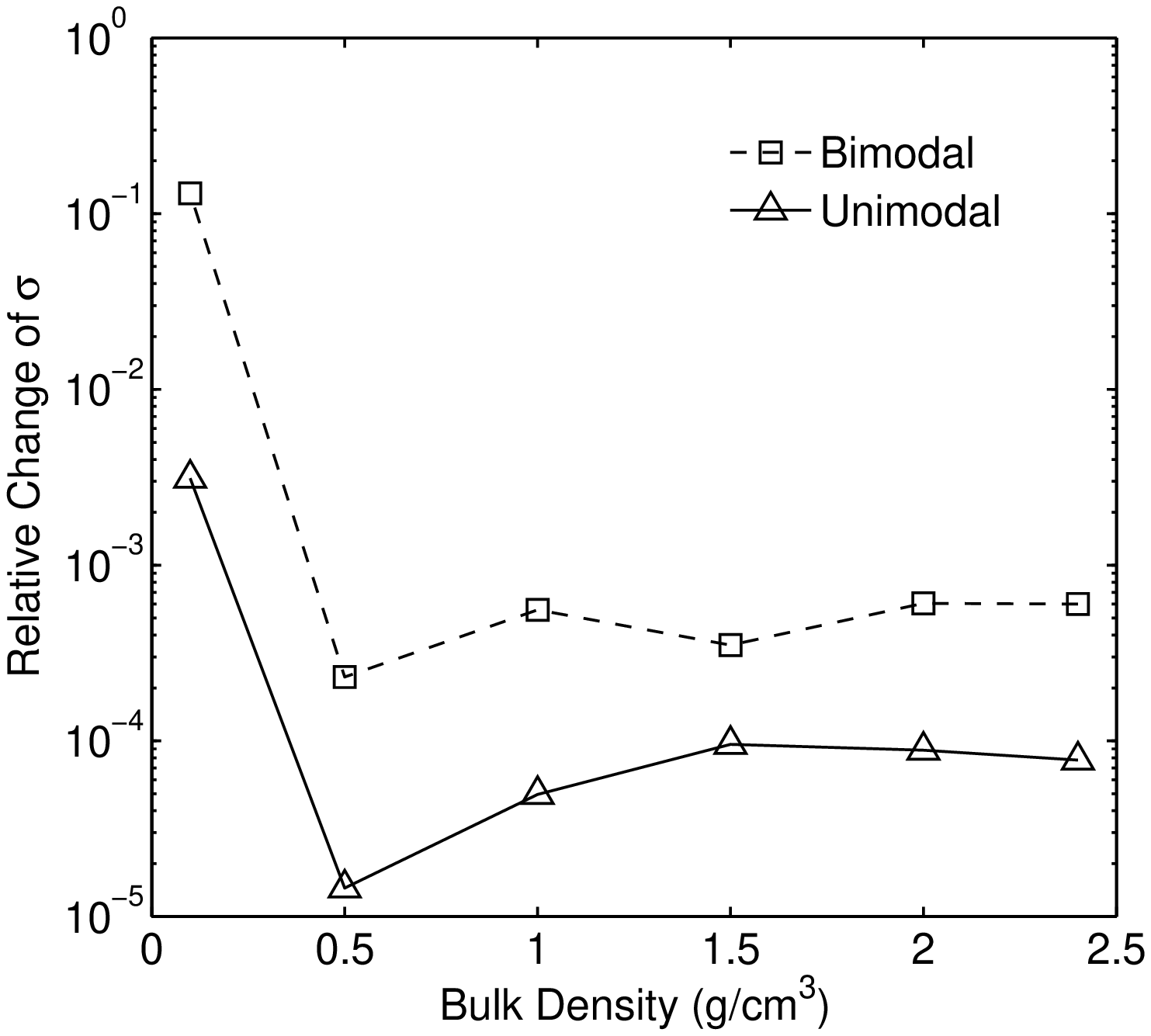} 
 } 
\subfigure[Rubble piles with ``smooth" parameter set.]{
   \label{fig:subfig3}
   \includegraphics[width=0.45\textwidth] {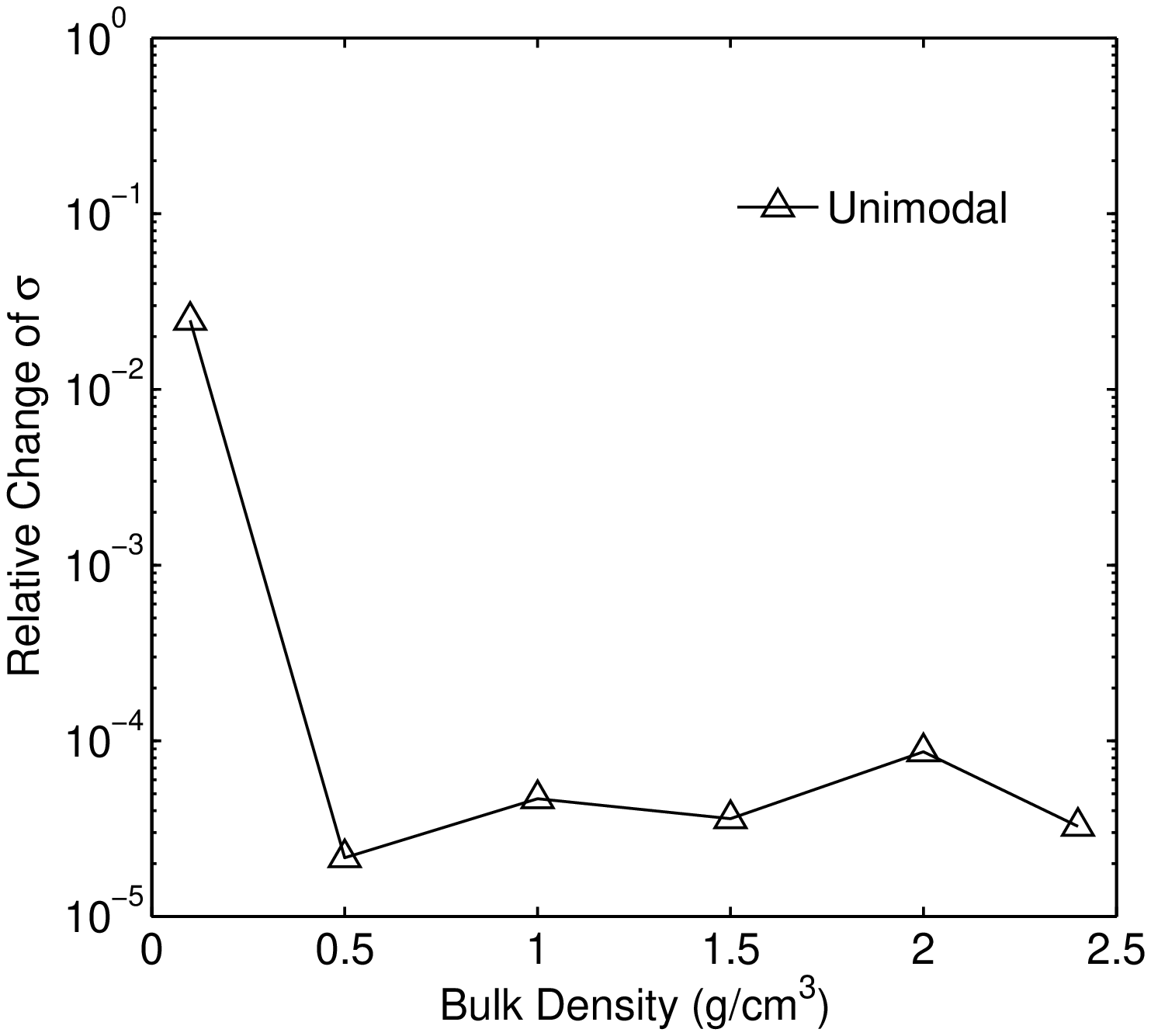} 
 }
\caption{}
\label{f:comp}
\end{figure}

\begin{figure}[h]
\centering
\scalebox{0.4}
{\includegraphics{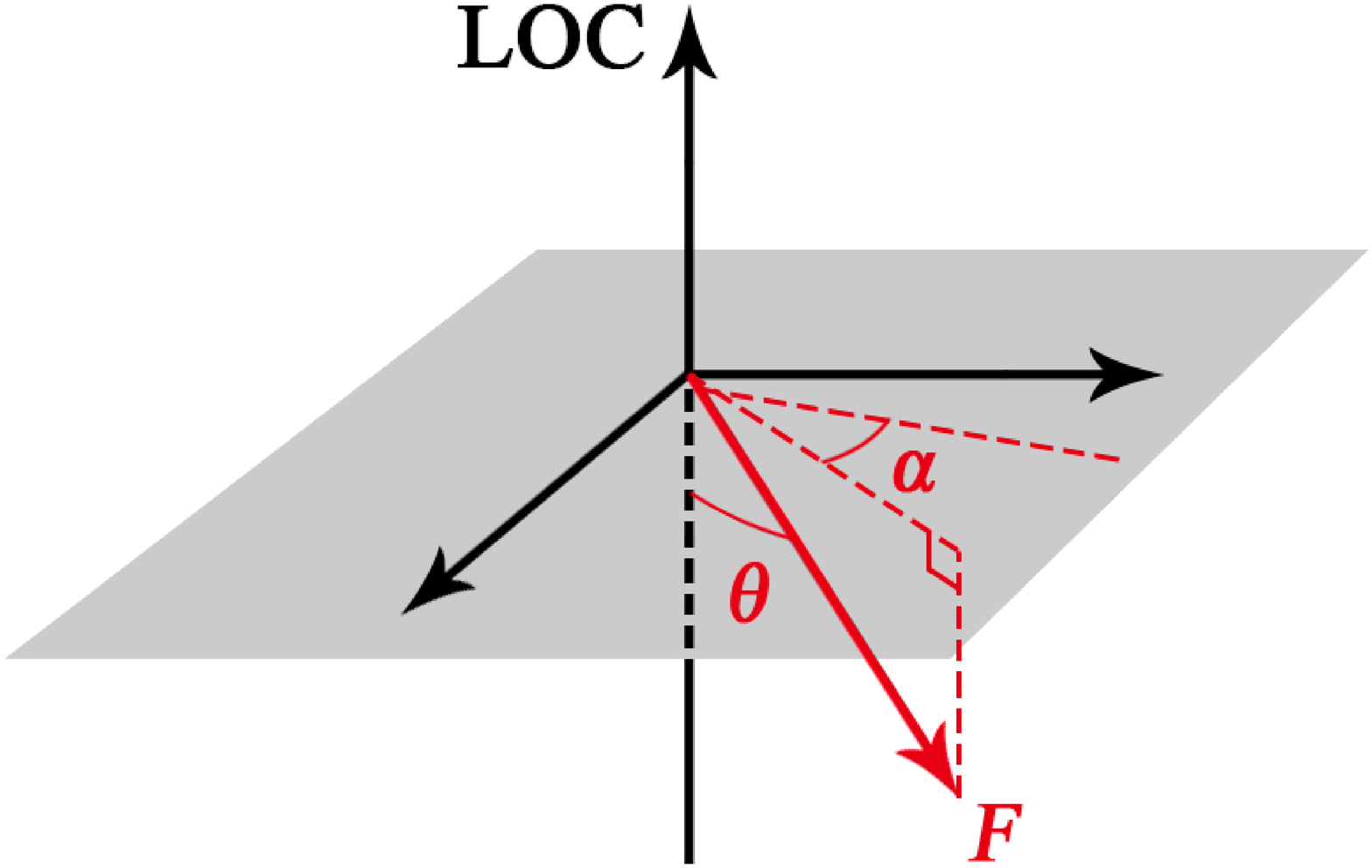}}
\caption{}
\label{f:ACCS}
\end{figure}

\begin{figure}[h]
\centering
\scalebox{0.45}
{\includegraphics{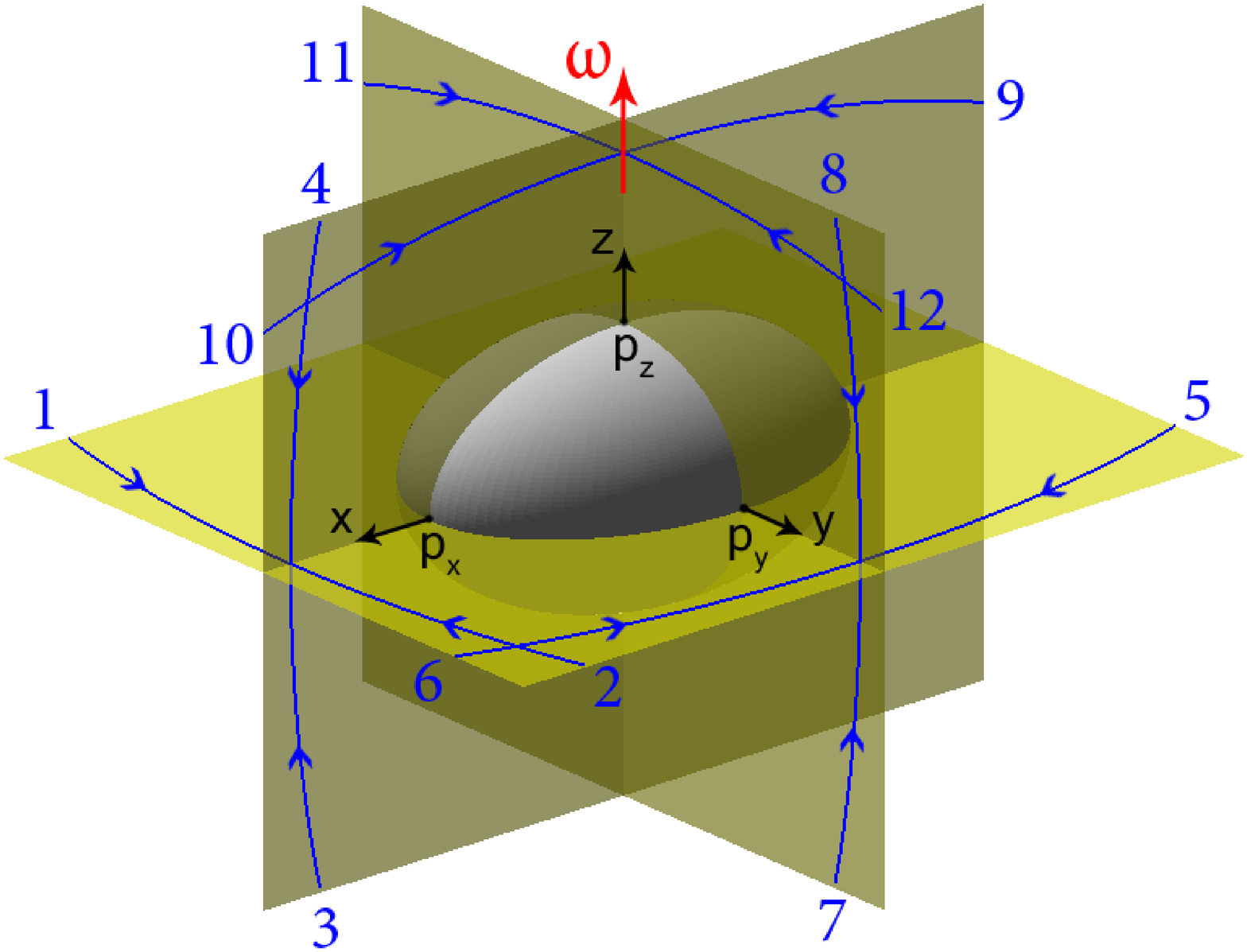}}
\caption{}
\label{f:traject}
\end{figure}

\begin{figure}[h]
\centering
\scalebox{0.7}
{\includegraphics{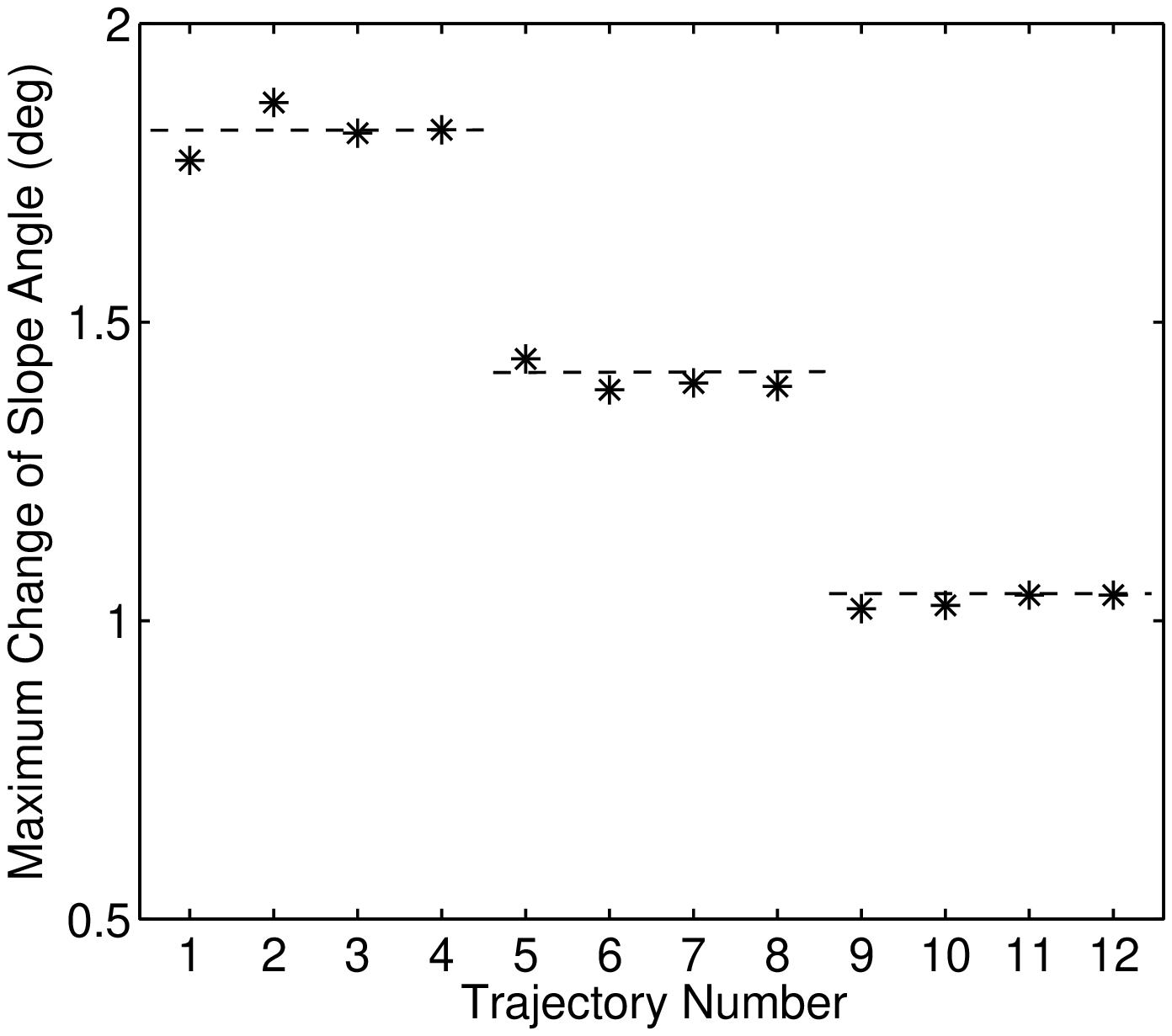}}
\caption{}
\label{f:maxCSA}
\end{figure}

\begin{figure}[h]
\centering
\scalebox{0.45}
{\includegraphics{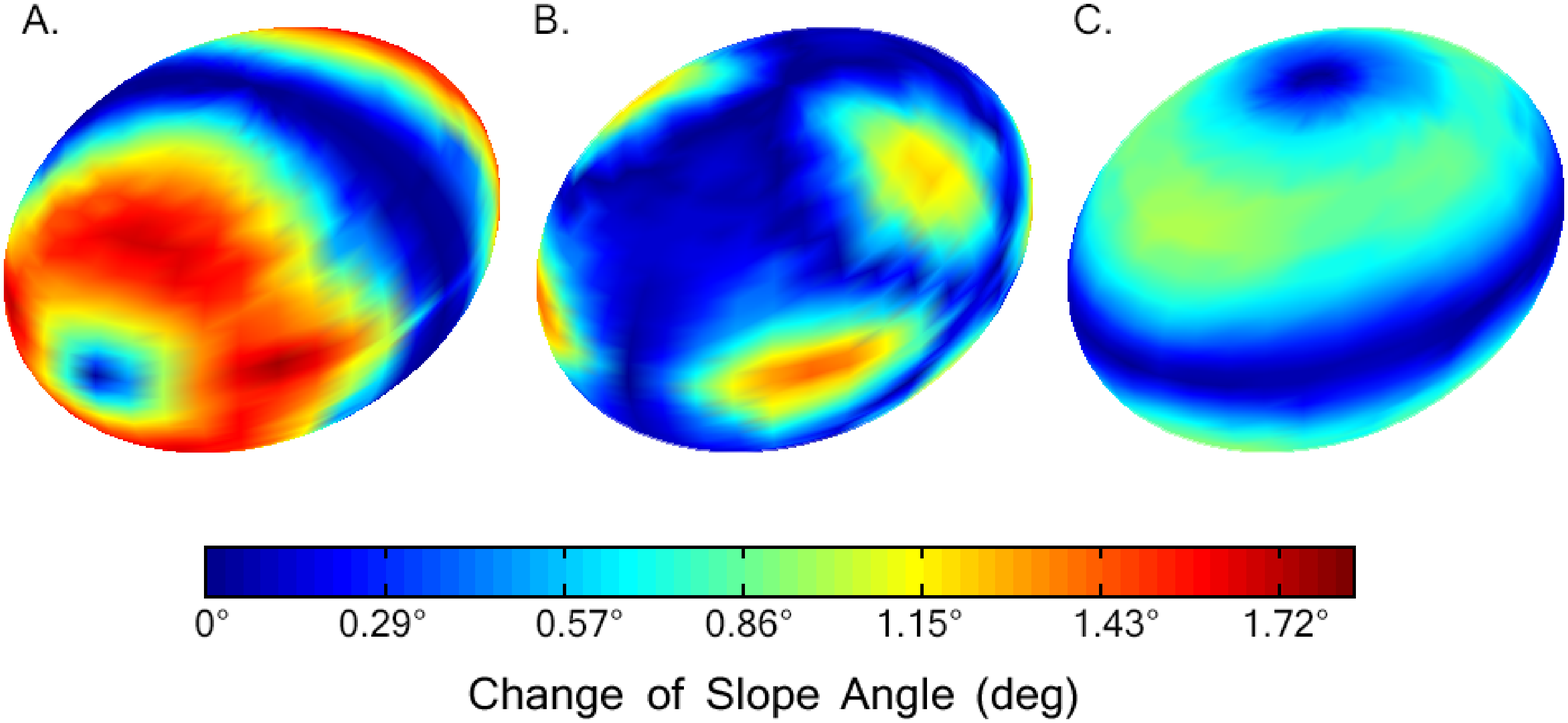}}
\caption{}
\label{f:patterns}
\end{figure}

\begin{figure}[h]
\centering
\scalebox{0.30}
{\includegraphics{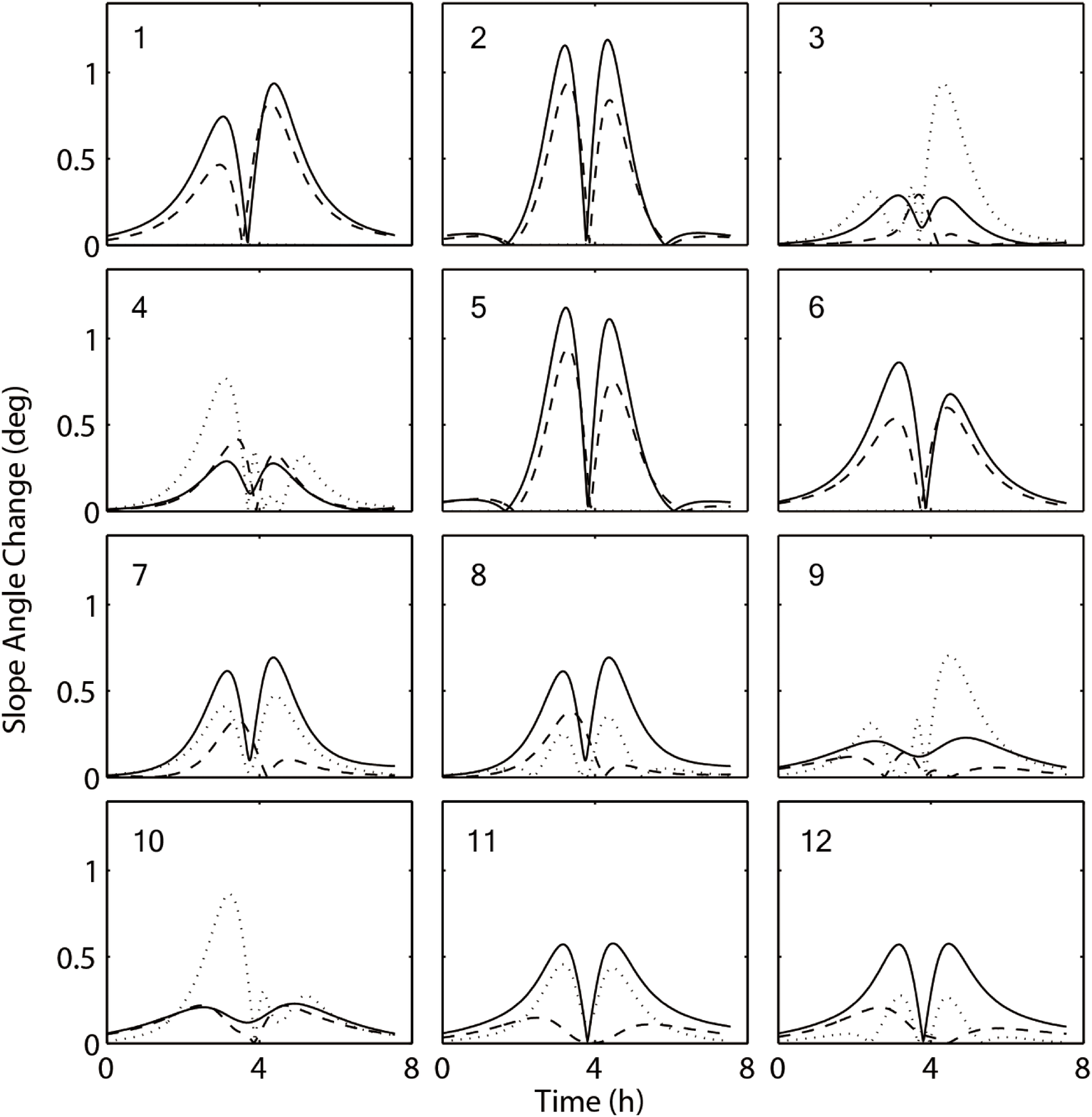}}
\caption{}
\label{f:pxpypz}
\end{figure}

\begin{figure}[ht!]
\centering
 \subfigure[Avalanches with ``gravel" parameter set.]{
   \label{fig:subfig1}
   \includegraphics[width=0.8\textwidth] {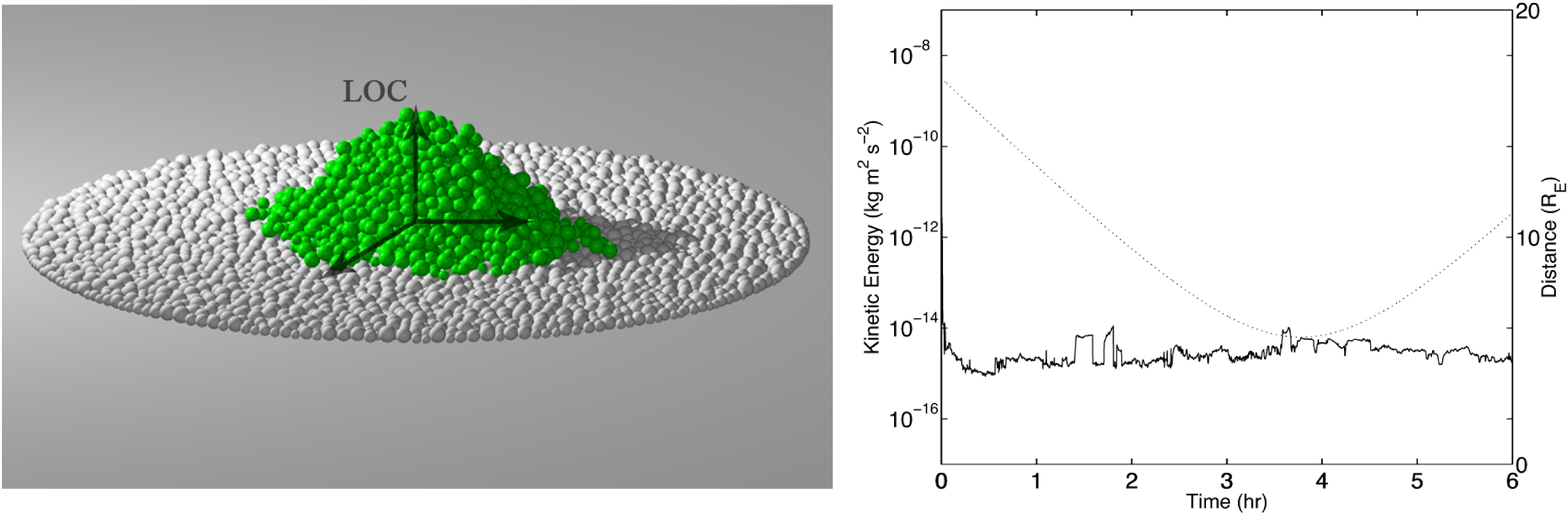}
 }
 \subfigure[Avalanches with ``glass beads" parameter set.]{
   \label{fig:subfig2}
   \includegraphics[width=0.8\textwidth] {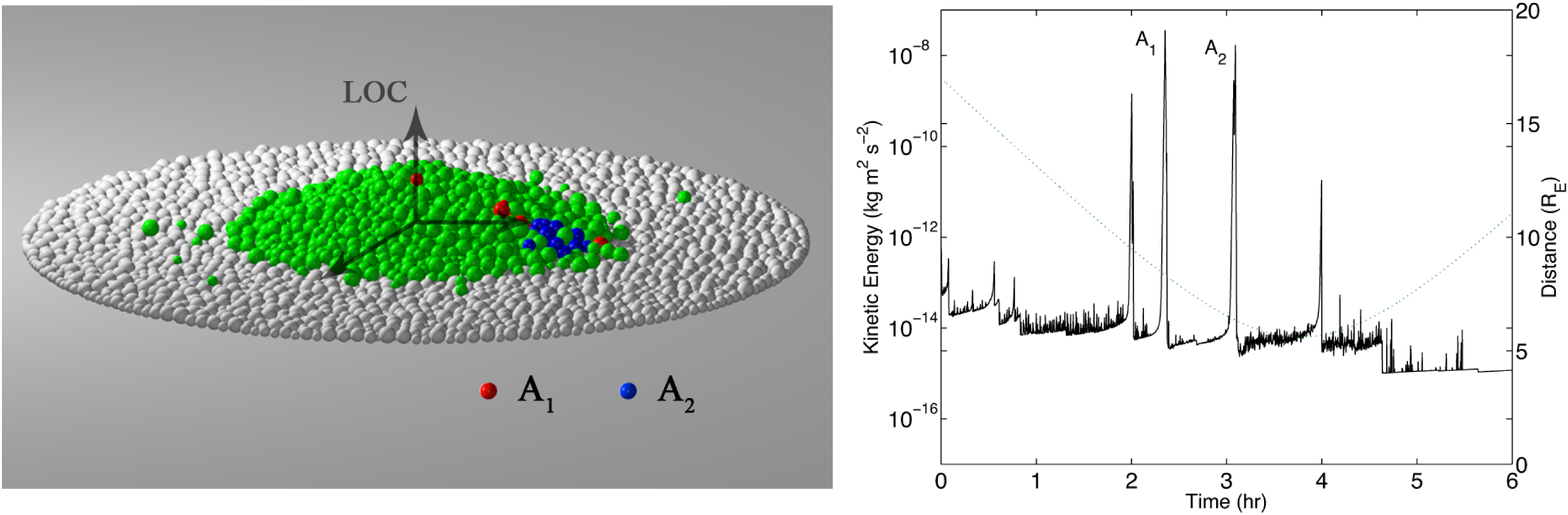} 
 }
 \subfigure[Avalanches with ``smooth" parameter set.]{
   \label{fig:subfig3}
   \includegraphics[width=0.8\textwidth] {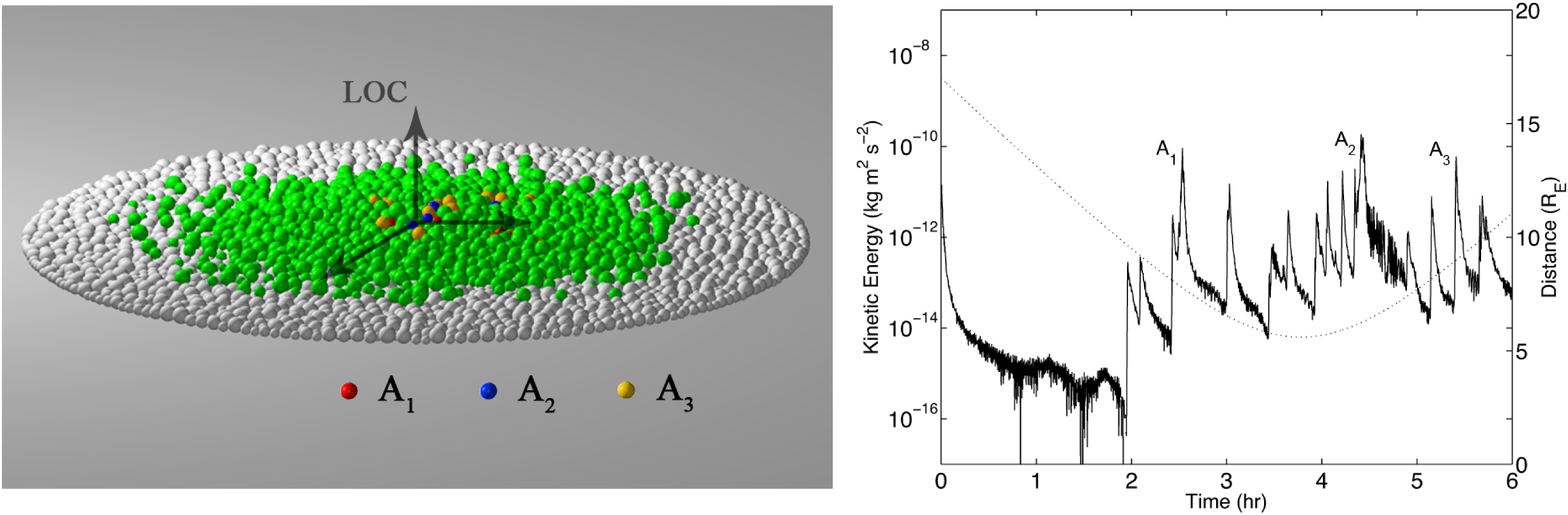}  
 } 
\caption{}
\label{f:avalanches}
\end{figure}

\begin{figure}[ht!]
\centering
 \subfigure[Shape changes with ``gravel" parameter set.]{
   \label{fig:subfig1}
   \includegraphics[width=0.85\textwidth] {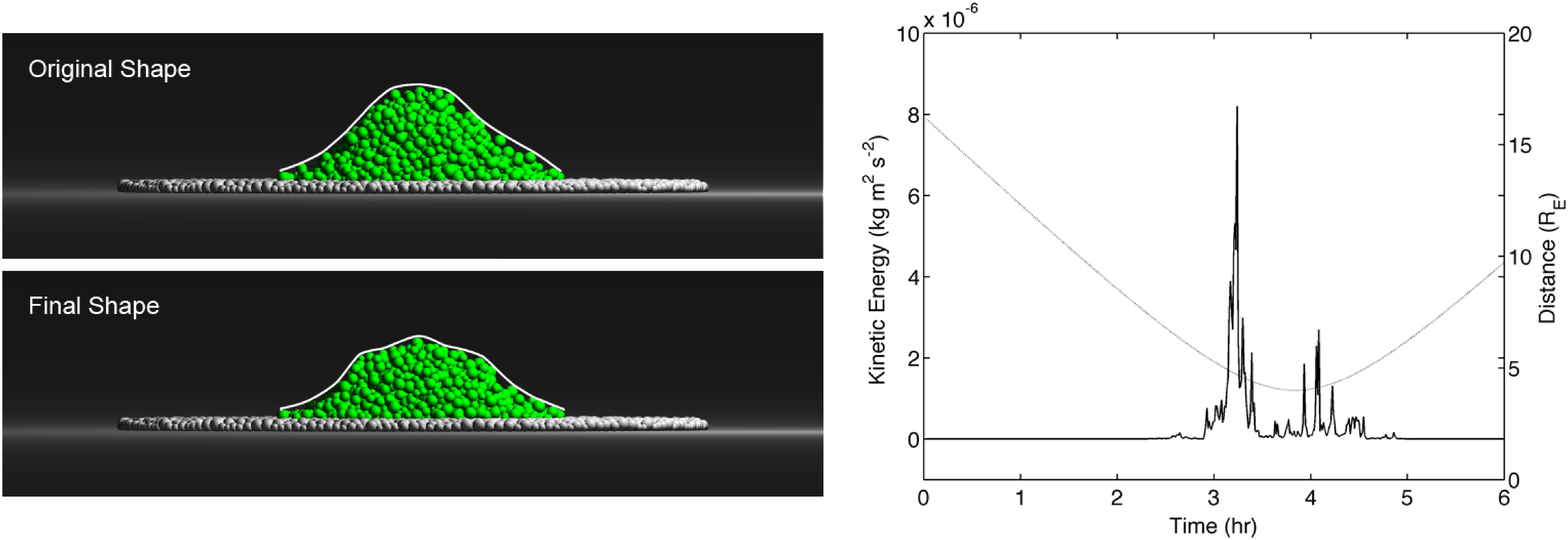}
 }
 \subfigure[Shape changes with ``glass beads" parameter set.]{
   \label{fig:subfig2}
   \includegraphics[width=0.85\textwidth] {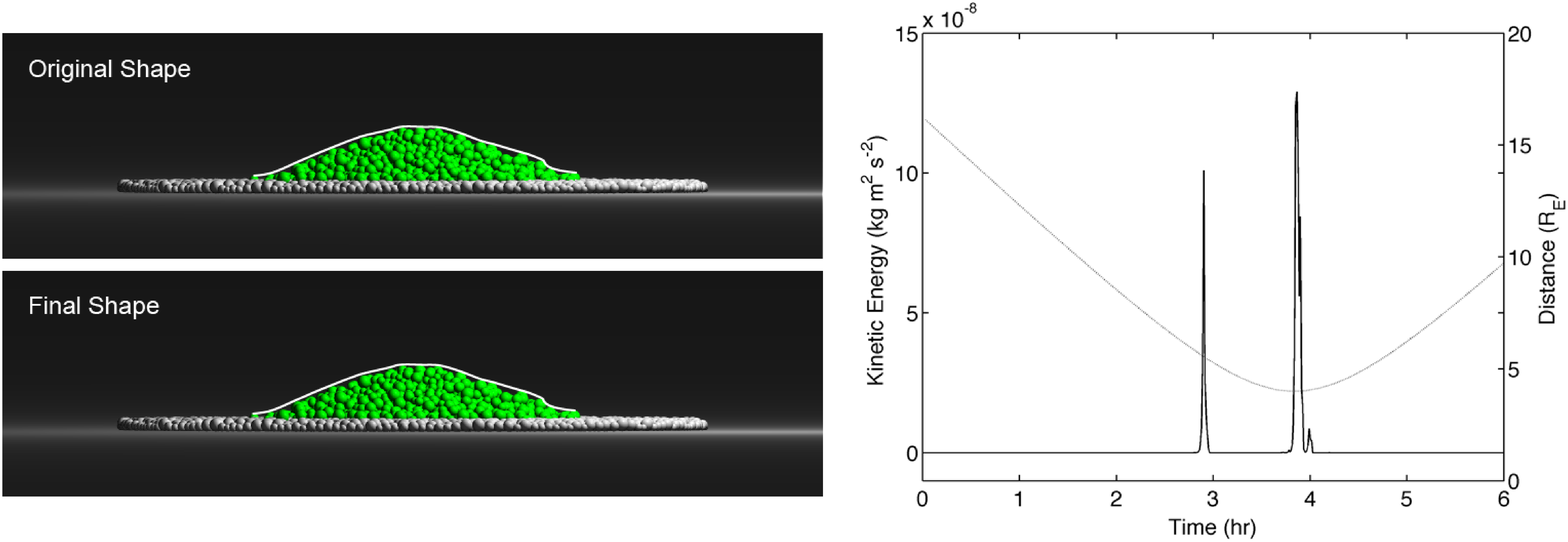} 
 }
 \subfigure[Shape changes with ``smooth" parameter set.]{
   \label{fig:subfig3}
   \includegraphics[width=0.85\textwidth] {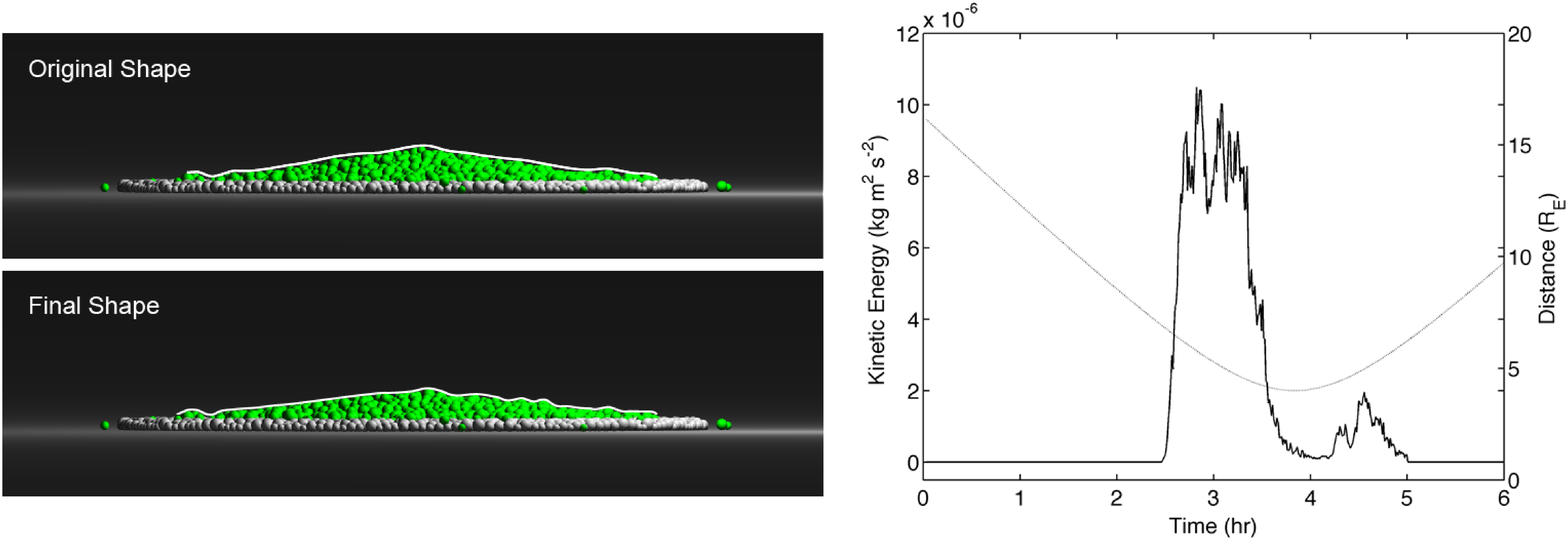}  
 } 
\caption{}
\label{f:refrsh4f}
\end{figure}

\begin{figure}[ht!]
\centering
 \subfigure[Shape changes with ``gravel" parameter set.]{
   \label{fig:subfig1}
   \includegraphics[width=0.85\textwidth] {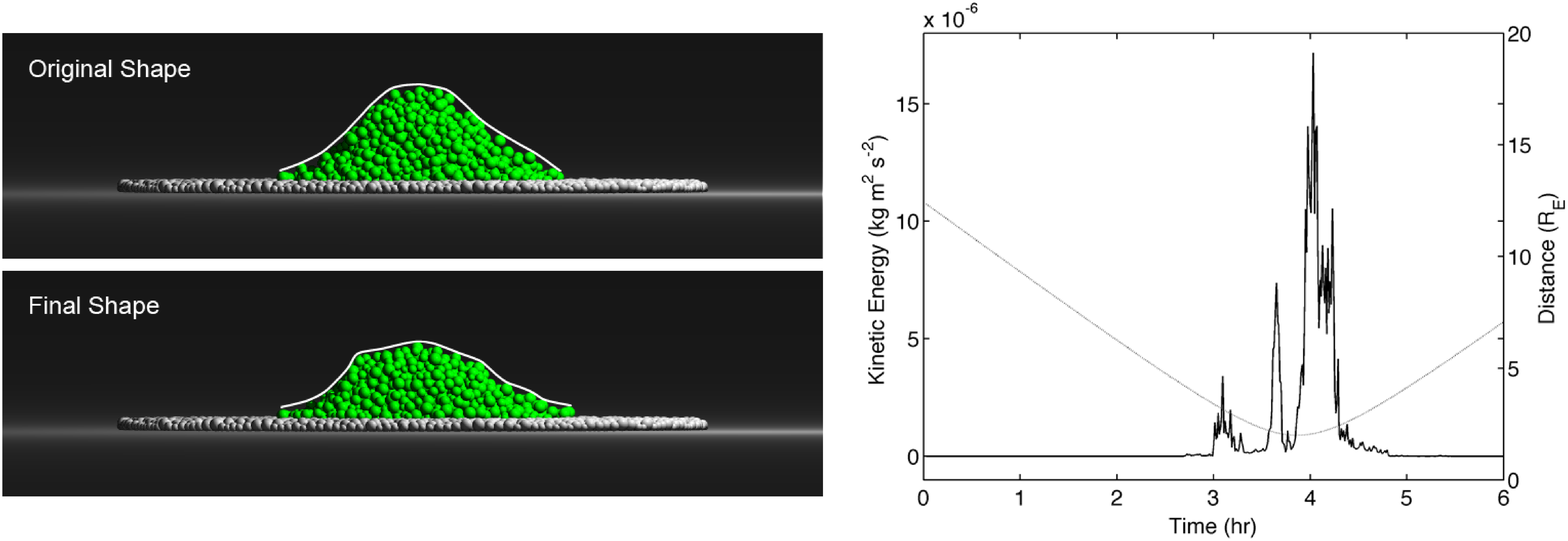}
 }
 \subfigure[Shape changes with ``glass beads" parameter set.]{
   \label{fig:subfig2}
   \includegraphics[width=0.85\textwidth] {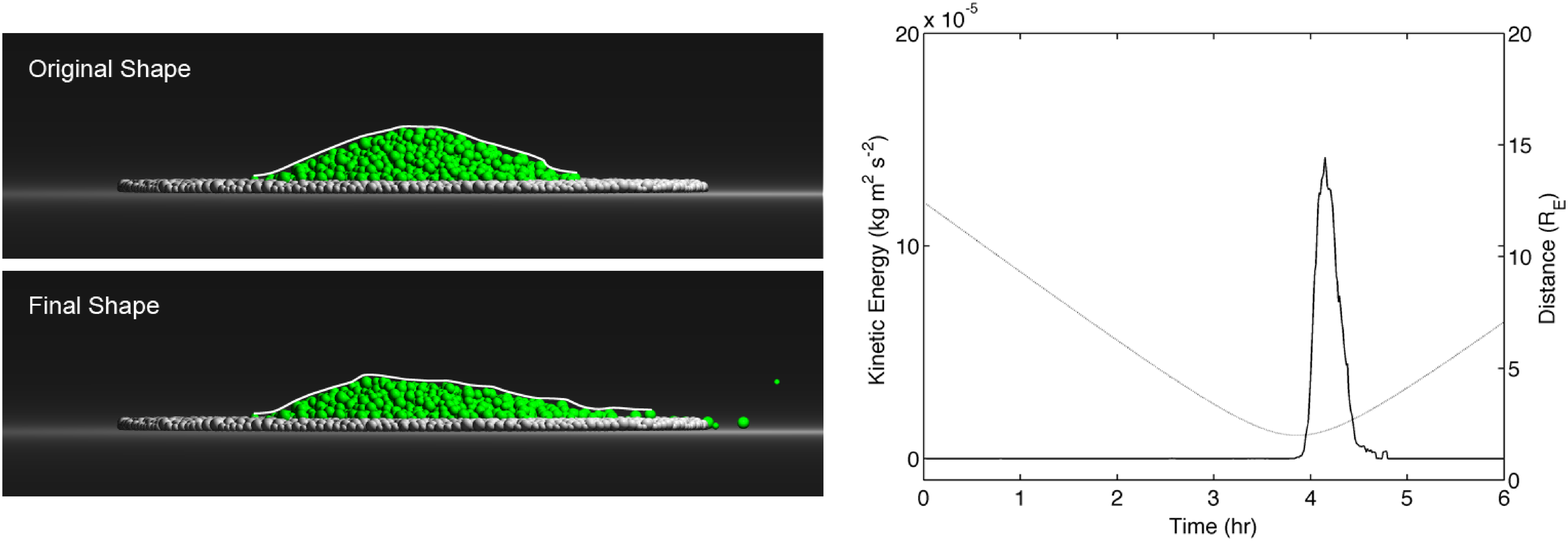} 
 }
 \subfigure[Shape changes with ``smooth" parameter set.]{
   \label{fig:subfig3}
   \includegraphics[width=0.85\textwidth] {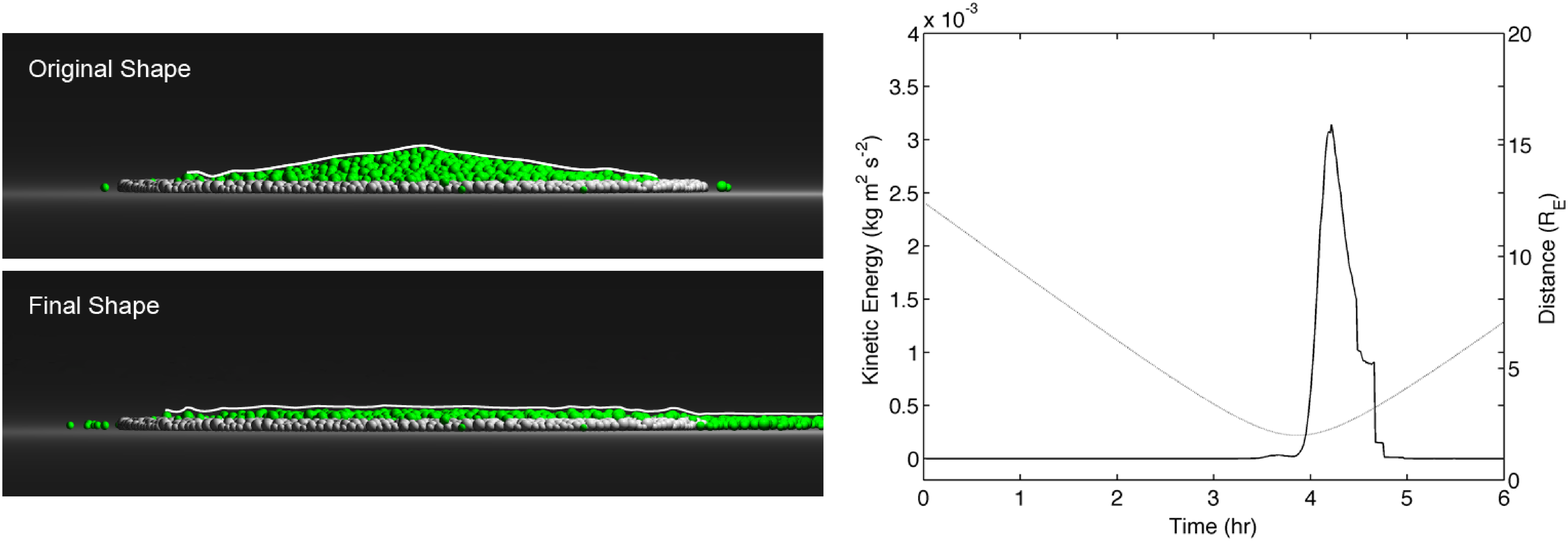}  
 } 
\caption{}
\label{f:refrsh2f}
\end{figure}

\begin{description}

  \figcap{sandpile}{ Snapshots of sandpiles constructed using three different materials, which are (a) ``gravel," (b) ``glass beads," and (c) ``smooth," with corresponding soft-sphere parameters listed in \tbl{SSDEMparams}. The same constituent spheres are used in each sandpile both for the free particles (green) and the rigid particles (white). The values of average slope and pile height after equilibrium is achieved are indicated in the snapshots.}

  \figcap{rockslide}{ Snapshots of the avalanche experiment and corresponding simulations. In the experiments, $\sim 340$ similar-size gravel pieces were piled up on the slope each time and released all at once by removing the supporting board. Snapshots from both the experiment and simulation include frames from the beginning (a, d), middle (b, e), and end (c, f) of the avalanche event, respectively.}

  \figcap{frames}{ Sketch of the frames used for deriving the motion equations of the local sandpile, indicated with different colors. SPC (black) is an inertial frame with the origin and axes set by  \code{pkdgrav}. MCT (blue) is a noninertial translating frame with the origin at the mass center of the asteroid and $x$, $y$, $z$-axes parallel to SPC.\ BDY (green) is a noninertial frame that is fixed to the asteroid and initially coincides with MCT.\ LOC (red) is a frame fixed to the asteroid surface, taking the origin to be where the sandpile is located and choosing the $z$-axis to be normal to the surface at that point. }

  \figcap{rubblepile}{ Diagram of Asteroid (99942) Apophis'. The tri-axial ellipsoid (light gray) denotes the overall shape model, while the arranged particles (khaki) denote the rubble-pile model used for numerical simulations. The highlighted particles (red) are markers used for determining the attitude of the asteroid in SPC. }

  \figcap{newpile}{ Snapshot of the rubble-pile model for Apophis with bimodal particles in irregular packing. The tri-axial ellipsoid (light gray) denotes the overall shape model, and the arranged particles (khaki) show the rubble-pile model used for numerical simulations. }

 \figcap{comp}{ The relative change of shape factor $\sigma$ as a function of bulk density during the encounter. The solid line (with triangular points) shows the results for the original rubble pile with equal-sized particles. The dashed line (with square points) shows the results for the bimodal rubble pile with irregular packing. }

  \figcap{ACCS}{ Sketch of the environmental force that the sandpile feels in LOC. The plane (gray) indicates the local tangential plane. The red arrow indicates the environmental force given by \eqn{EnvirForce}. Label $F$ indicates its magnitude, and $\theta$ and $\alpha$ indicate the slope angle (elevation) and deflection angle (azimuth) of the effective gravity, respectively. }

  \figcap{traject}{ Sketch of the fiducial trajectories in different orientations. These trajectories (blue numbered lines) are placed in the perpendicular planes (yellow) of the Apophis model (gray ellipsoid), matching the three axial directions at perigee. The blue arrows show the prograde and retrograde directions. The red arrow denotes the angular velocity. p$_x$, p$_y$, p$_z$ indicate the poles along the three axes $x$, $y$, $z$, respectively. }

  \figcap{maxCSA}{ Maximum values of slope angle change for the fiducial trajectories. The stars denote the maximum values, and the dashed lines denote the levels of tidal perturbation  (the average of the 4 values in each case), which divide the $12$ trajectories into $3$ basic catagories according to the spin orientation at perigee. See \fig{traject} for the trajectory orientations. }

  \figcap{patterns}{ Global distribution of slope angle change at perigee for the three trajectory sets A, B and C, each including 4 trajectories of the same spin orientation at perigee (see text for an explanation of each set). A uniform colormap ranging up to $1.8^\circ$ is used for the three plots. }

  \figcap{pxpypz}{ Time variation of the effective gravity slope angle at the poles p$_x$ (dashed lines), p$_y$ (solid lines), and p$_z$ (dotted lines) during the encounter. Each subgraph corresponds to one fiducial trajectory marked with its number on the upper left (refer to \fig{traject}). }

  \figcap{avalanches}{ Snapshots of sandpiles after the tide-induced avalanches in 2029, with corresponding time variations of their total kinetic energy. The three panels include the results of sandpiles constructed using the three materials (a) ``gravel," (b) ``glass beads," and (c) ``smooth," respectively. Particles that moved more than $0.1$ of a particle radius during the avalanches are highlighted. The peaks in the total kinetic energy curves indicate occurrence of avalanches; the peaks are labeled, and corresponding particles that moved appreciably are marked in different colors (red, blue, orange) in the snapshots. The dotted line in each graph shows the evolution of the planet-asteroid distance during the encounter.}
 
  \figcap{refrsh4f}{ Snapshots of the sample sandpiles before and after the avalanches induced by the encounter at $4.0$ $\mathrm{R_E}$ perigee, with corresponding time variations of their total kinetic energy. The white lines in the snapshots indicate the original and final shapes of the sandpiles. The peaks in the total kinetic energy curves indicate occurrence of avalanches, and the dotted line shows the variation of the planet-asteroid distance during the encounter.}

  \figcap{refrsh2f}{ Same as \fig{refrsh4f}, but for the encounter at $2.0$ $\mathrm{R_E}$ perigee.}

\end{description}

%
%

\clearpage

\clearpage

\clearpage

\clearpage

\clearpage

\clearpage

\clearpage

\clearpage

\clearpage

\clearpage

\clearpage

\clearpage

\clearpage

\clearpage

\end{document}